\documentclass[floats,twocolumn,aps,prb,longbibliography,superscriptaddress]{revtex4-2}

\usepackage[export]{adjustbox}
\usepackage{graphicx}
\usepackage{dcolumn}
\usepackage{bm}
\usepackage{physics}
\usepackage{float}
\usepackage{subfig}
\usepackage{color}
\usepackage{array}
\usepackage{makecell}
\usepackage{ragged2e}
\usepackage[font=small,
   justification=centerlast, singlelinecheck=off, textformat = simple, format = plain]{caption}

\usepackage{bbold}

\begin{document}
	

	\title{Bound states and deconfined  spinons in the dynamical structure factor of the $J_1 - J_2$ spin-1 chain }

	\author{Aman Sharma}
	\affiliation{Institute of Physics, \'Ecole Polytechnique F\'ed\'erale de Lausanne, CH-1015 Lausanne, Switzerland\looseness=-1}
	
	\author{Mithilesh Nayak}
	\affiliation{Institute of Physics, \'Ecole Polytechnique F\'ed\'erale de Lausanne, CH-1015 Lausanne, Switzerland\looseness=-1}
	\affiliation{Department of Physics, University of Fribourg, 1700 Fribourg, Switzerland\looseness=-1}
	\author{Henrik M. R{\o}nnow}
	\affiliation{Institute of Physics, \'Ecole Polytechnique F\'ed\'erale de Lausanne, CH-1015 Lausanne, Switzerland\looseness=-1}
	
	\author{Fr\'ed\'eric Mila}
	\affiliation{Institute of Physics, \'Ecole Polytechnique F\'ed\'erale de Lausanne, CH-1015 Lausanne, Switzerland\looseness=-1}

	\date{\today}

\begin{abstract}
Using a time-dependent density matrix renormalization group approach, we study the dynamical structure factor of the $J_1 - J_2$ spin-1 chain. As $J_2$ increases, the magnon mode develops incommensurability. The system undergoes a first-order transition at $J_2 = 0.76 J_1$, and at that point, domain walls lead to a continuum of fractional quasi-particles or spinons. By studying small variations in $J_2$ around the transition point, we observe the confinement of spinons into bound states in the spectral function and find a smooth evolution of the spectrum into magnon modes away from the phase transition. We employ the single-mode approximation to accurately account for the dispersion of the magnon mode away from the phase transition and describe the associated continua and bound states. We extend the single-mode approximation to describe the dispersion of a spinon at the phase transition point and obtain its dispersion throughout the Brillouin zone. This allows us to relate the incommensurability at and around the transition point to the competition between a negative nearest-neighbor hopping amplitude and a positive next-nearest-neighbor one for the domain wall.
\end{abstract}

\maketitle

\section{Introduction}

Frustrated magnets are a platform to realize various exotic phenomena arising from competing couplings \cite{lacroix2011introduction}. The $J_1 - J_2$ antiferromagnetic spin chain is a paradigmatic model that embodies multiple fundamental aspects of quantum many-body physics, including quantum criticality, fractionalization, and topological order. The Hamiltonian for the $J_1 - J_2$ model is given by:
\begin{equation}
H = J_1 \sum_{i} \mathbf{S}_i \cdot \mathbf{S}_{i+1} + J_2 \sum_{i} \mathbf{S}_i \cdot \mathbf{S}_{i+2},
\label{Eq_Hamiltonian}
\end{equation}
where $\mathbf{S}_i$ represents spin operators. The behavior of this model varies significantly depending on the spin quantum number and the relative strengths of nearest-neighbor ($J_1$) and next-nearest-neighbor ($J_2$) interactions. In the classical limit, frustration results in incommensurate correlations when $J_2/J_1 > 0.25$ \cite{villain1959structure}. 

In the quantum version, the spin-1/2 chain exhibits incommensurability beyond $J_2/J_1 = 0.5$ (also known as the Majumdar-Ghosh point) \cite{bursill1995numerical,nomura2003onset}. If $J_2 = 0$, the model corresponds to a simple unfrustrated Heisenberg chain where the elementary excitations are fractionalized quasi-particles called spinons, which were first described exactly by the Bethe ansatz \cite{bethe1931theorie, clozieaux1962spin}. The simple model of Eq.~\ref{Eq_Hamiltonian} encapsulates two key phenomena: fractionalization and incommensurability. For the $J_1 - J_2$ spin-1/2 chain, a phase transition occurs at $J_2/J_1 = 0.2411$, separating a gapless phase from the gapped dimerized phase \cite{tonegawa1987ground, okamoto1992fluid, eggert1996numerical}. The ground state in the dimerized phase is doubly degenerate, characterized by alternating nearest-neighbor bonds with strong and weak spin-spin correlations. In either phase, magnetic excitations create pairs of domain walls or spinons that propagate freely, forming a two-spinon continuum. However, in the dimerized phase this does not fully describe the two-spinon spectrum, there also exist bound states of spinons. The spinon dispersion is a sum of cosines with a minimum at $k = \pi/2$. As $J_2/J_1$ increases beyond $\sim0.53$, the spinon dispersion becomes incommensurate. \cite{caspers1982some, lavarelo2014spinon}.

In contrast, the spin-1 chain exhibits qualitatively different physics. For $J_2 = 0$, the system is in the Haldane phase \cite{haldane1983continuum}, a gapped phase whose ground state can be described by a valence bond solid (VBS) consisting of singlets between the nearest neighbors \cite{aklt1987rigorous,auerbach1998interacting}. The system remains gapped for all values of $J_2$. At $J_2 \approx 0.75 J_1$, a first-order phase transition occurs, leading to a next-nearest-neighbor (NNN) Haldane phase, where singlets form between the next-nearest neighbors \cite{kolezhuk1996first, pixley2014frustration, chepiga2016dimerization}. At the transition point, the lowest-energy excitations are spinons, domain walls separating regions of Haldane and NNN Haldane ground states. In this regime, the spinon dispersion has been calculated using a quasi-particle ansatz. These spinons have an incommensurate dispersion with a small gap of approximately $0.015 J_1$\cite{vanderstraeten2020spinon}. Beyond the transition, the spinons experience an effective confining potential, giving rise to multiple bound states. Far from the transition point, in either the Haldane or NNN Haldane phase, the lowest excitations are spin-1 excitations. 

Although the low-energy spectrum has been studied\cite{vanderstraeten2020spinon}, a comprehensive understanding of the connection between spinon-based excitations near the phase transition and magnon-based excitations far from the transition is lacking. In this work, we fill this gap by studying the dynamical structure factor using a time-dependent density matrix renormalization group (tDMRG) method. We demonstrate how spinons evolve into magnons as $J_2$ increases, and we explore how incommensurability arises from the competition between the nearest-neighbor and next-nearest-neighbor interactions.
 Furthermore, the discussion of the spectrum in the literature has been mostly about the low-lying excitations and a description of possible high-energy excitations is lacking. In this paper, we provide a numerical investigation of the spectrum in the full energy range by computing the dynamical structure factor (DSF) using t-DMRG algorithm for a range of $J_2$ values. In order to interpret the spectral features we determined the dispersion of elementary excitations i.e.  magnons or spinons, by using the single mode approximation (SMA). Multi-particle continua are built starting from these elementary excitations, and various modes which lie outside of the continua are identified to be bound states.  The power of the SMA analysis to describe incommensurability has been demonstrated in earlier works on the spin-1 $J_1-J_2$ chains and bilinear biquadratic chains \cite{kolezhuk1997variational}. Here also the SMA analysis gives further insights into the origin of incommensurability in the spinon dispersion. Since the DSF can be directly measured for a real compound in inelastic neutron scattering (INS) experiments, our study can be of relevance to future experimental studies. \par 

The paper is organized as follows. In section \ref{methods}, we briefly describe the t-DMRG method used to obtain the DSF for the $J_1 - J_2$ spin-1 chains and the SMA approach to obtain the dispersion of the prominent magnon mode or a single domain wall.  In section \ref{numerical_spectral_functions_J1_J2}, we describe the DSF results and identify various modes in the rich spectra. We describe the following interesting phenomena - (i) the single magnon mode develops incommensurability as $J_2$ is increased and is explained using SMA, (ii) there exist bound states of magnons which lie outside of the multi-magnon continua, and (iii) close to the transition point the spinons are deconfined, resulting in a continuum. We determine the spinon dispersion at the transition point throughout the Brillouin zone inspired by the SMA analysis with parameters obtained by fitting with the low energy dispersion from Ref.\cite{vanderstraeten2020spinon}  and our numerical DSF at the transition point. Finally, we give an outlook by discussing interesting extensions for the DSF of larger-spin chains and the possible experimental observation of some of these rich spectra.	
	
\section{Methods}
\label{methods}
\subsection{Time-dependent DMRG}	
\label{tDMRG_benchmarking}

The dynamical structure factor (DSF) of a system at zero temperatures is given by: 
\begin{eqnarray}
S^{\alpha, \alpha}(k,\omega)=\int dt e^{-i \omega t}\sum_{r_i,r_j}e^{i k(r_i-r_j)} \bra{\psi_0}S^\alpha_{r_i} (t) S^\alpha_{r_j} \ket{\psi_0},\nonumber
\end{eqnarray}
where $\alpha= x, y, $ or $z$ and $|\psi_0\rangle$ is the groundstate of the system. The DSF is simply the double Fourier transform of the time-dependent spin-spin correlation functions of the system. It is proportional to the differential-scattering cross section in an INS experiment. In principle, by computing spin correlations starting from different spin operators, one effectively determines various components of the differential scattering cross-section. Since there is no anisotropy in the system being considered, we primarily focus our investigation of the spectra by computing the longitudinal component of the DSF (i.e. $S^{zz}(k,\omega)$). 

 It is evident in the formulation that one needs to follow a two-step procedure to compute the time-dependent correlations - (i) obtain the ground state of a 1D chain, and (ii) evolve the state resulting from application of the local spin operator to the ground state. DMRG has been very successful in obtaining the ground state of a 1D model and its properties \cite{white1992density, white1993density, schollwock2011density}.  In order to converge to a groundstate for a finite chain of 200 sites, we kept bond dimension of 250 states in the matrix-product-states (MPS)-ansatz and sweeped through the system with one-site DMRG till the simulation reached a variance per site of $10^{-8}$. This variance was achieved in spin-1 chains for $J_2<0.6 J_1$ but it became increasingly tough to converge with the variance for larger values of $J_2$. Therefore, we relaxed the convergence criterion for variance to be $10^{-5}$ or less.\par 
 
 The time evolution of the state resulting from applying the spin-operator onto the ground state is then carried out by the t-DMRG method \cite{white2004real, white2008spectral}. The time evolution operator $U (t) = \mathrm{exp}(-it H)$ is exponential in dimension owing to the exponentially large Hilbert space of the Hamiltonian. The method proceeds by generating a set of Trotterized evolution gates. These gates evolve the system by a small time-step size when applied to a state. 
  For the simulation results obtained on the $J_1 - J_2$ spin-1 chain,  we employed first-order Trotter gates with very small time steps of  $0.02/J_1$ for  values of $J_2$ less than $0.8 J_1$. For larger $J_2$ couplings, we used a smaller time-step of  $0.01/J_1$. The application of individual gates to the adjacent MPS increases the bond dimension which is then truncated to a target bond dimension so that one does not end up storing an exponential number of states. We fixed the target bond dimension during time evolution to be same as that used for the ground state, namely 250 states.\par
  
     Applying a local spin operator to the center of the chain leads to a state which is a superposition of some allowed excited eigenstates of the system. The time-evolution of the system effectively propagates this localized perturbation across the chain. Propagating the localized perturbation results in increasing the number of allowed states in the MPS-ansatz. A linear growth of entanglement entropy leads to exponentially large MPSs, which is taken care of by truncating the evolved MPS to the target bond dimension. The propagation of the perturbations in the chain also allows the spin-spin correlations to spread in the chain. While simulating on a finite chain, one should take care of not letting the correlations reach the boundaries of the chain. In order to avoid the artefacts from the boundaries influencing the correlation function, we performed the time evolution of the state until the time at which the correlations reached the lattice boundaries.  We note that the velocity associated with the spread of correlations depends upon the value of $J_2$, and since the length of the chain is fixed to 200 sites, we could evolve the system up to a final time that depends on the value of $J_2$ (see Fig.\ref{Fig_appendix_velocities} in the appendix). \par
     
      One obtains the DSF ($S^{z,z}(k,\omega)$) from a double Fourier transform of the time-dependent spin-spin correlation functions in space and time. Since the groundstate and time-evolution calculations were done with open boundary conditions, one can choose momentum values continuously.  Since one can evolve the finite chains for finite time only, the Fourier transform in time will result in ringing effects in the DSF. This is suppressed by convoluting the time-dependent correlations with a Gaussian function $\sim e^{-\frac{t^2}{2\sigma^2}}$ . Naturally this filter broadens the spectral function leading to loss of resolution in the spectra. In order to optimize the resolution of the modes without suffering from the artefacts, we chose the variance ($\sigma$) to be $0.276~t_f$ in the current simulations.\par
  
\begin{figure}
\centering
\includegraphics[height = 11cm, width = 11cm, keepaspectratio ]{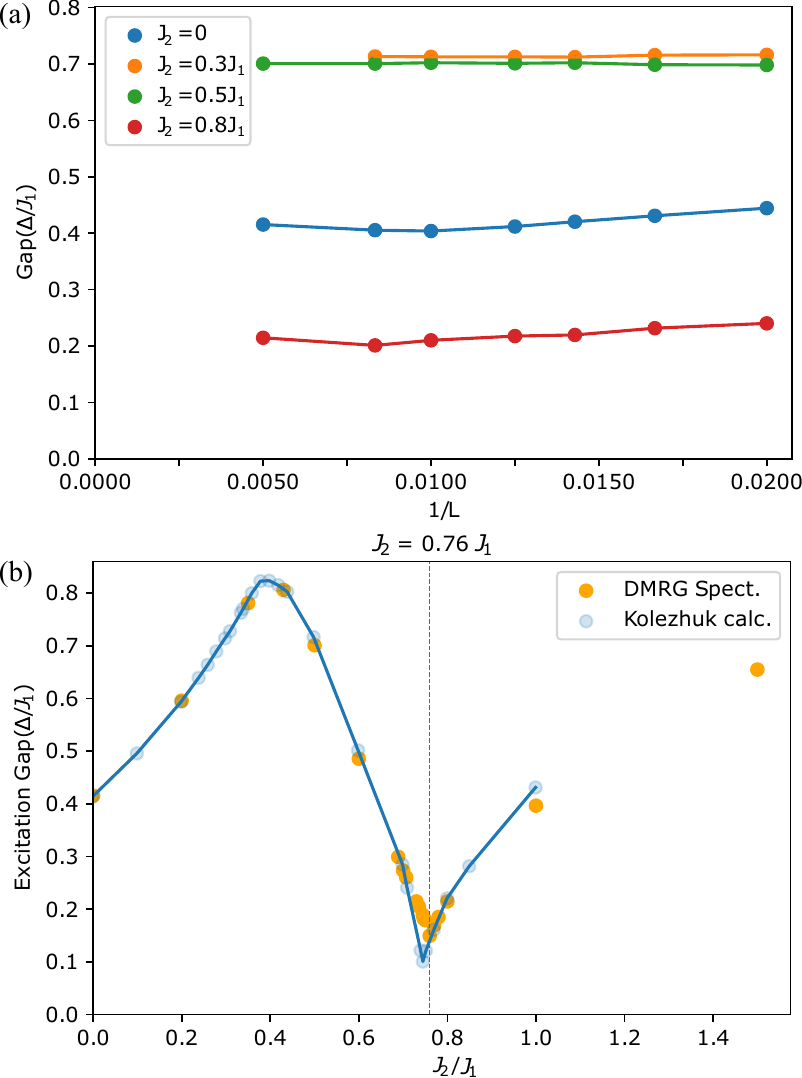}
\caption{(a) Variation of the excitation gap in the numerical DSF of the $J_1 - J_2$ spin-1 chain with the length of the chain for different $J_2$ values. (b) Comparison of the energy difference between the first excited state and the groundstate (blue dots) with the spectral gap in DSF (orange dots). The energy difference data is taken from Ref. \cite{kolezhuk1996first} which used DMRG groundstate simulations to find the gap and have been scaled in length for all values of $J_2$. The spectral gap is obtained from measuring the position of the lowest energy excitations in the DSF only for chains of length 200 sites. One finds a drop in the spectral gap at the phase transition, which is determined to be at $J_2 = 0.76J_1$.} 
\label{Fig1}
\end{figure}

  The Lehmann representation of the longitudinal DSF is given by : 
  \begin{eqnarray}
   S^{zz}(k,\omega)= \frac{2\pi}{N}\sum_{\alpha} |\bra{\psi_\alpha} S^z_{-k}\ket{\psi_0}|^2\delta\left(\omega  - \omega_\alpha(k)\right),
   \label{Eq_Lehmann_rep}
  \end{eqnarray}    
where $N$ is the length of the chain and $\omega_\alpha(k)$ is the change in energy from the ground state $|\psi_0\rangle$ to the final state $|\psi_\alpha\rangle $. As the ground state is in the total spin-0 sector, the spin operator $S^z$ only connects to excitations in the spin-sector with total spin 0 or 1. Thus, the spectral gap seen in the DSF is due to the excitation in the spin-1 sector.  The gap can be exactly calculated using DMRG since the first excited state is the lowest energy state of the $S=1$ sector and therefore one can simply find the groundstate after screening the edge states from running the DMRG code only in this symmetry sector.  We compare the spectral gap in the DSF with the DMRG determined gap taken from Refs \cite{kolezhuk1996first, kolezhuk1997variational, kolezhuk2002connectivity}  in Table \ref{Table1_benchmarks}. We measured the spectral gap in the DSF to be the lowest energy spectral peak observed in the DSF. This was carried out on finite chains with varying lengths for selected $J_2$ values and we obtain a nearly size independent gap (see Fig. \ref{Fig1} a). 
As $J_2$ is tuned closer to the phase transition point, the correlation length increases and the spectral gap becomes too small to be accurately determined from the DSF data for finite chains. The Gaussian filter on the closely spaced excitation levels observed in the DSF,  near the transition point, results in an overestimation of the spectral gap. We highlight the dependence of the spectral gap on $J_2$ by plotting the spectral gap for a fixed-length chain  against the gap in excitations determined by Kolezhuk {\it et.al.} \cite{kolezhuk1996first} in Fig. \ref{Fig1} b. Apart from the immediate vicinity of the transition point, where our gap is larger than that of Ref. \cite{kolezhuk1996first} due to finite size effects, the agreement is excellent. Note that the actual gap at the transition is even much smaller than that of Ref. \cite{kolezhuk1996first} , as demonstrated by the spinon calculation of Ref.\cite{vanderstraeten2020spinon} (see discussion below).

\begin{table}
\centering
\caption{Numerical comparison of energy difference between the first excited state and the groundstate energy as calculated in Ref. \cite{kolezhuk1996first} denoted by $E_1- E_0$ and the spectral gap values obtained from DSF denoted by $\Delta E$.}
\begin{tabular}{||c c c c||} 
\hline
$J_2$ & $E_1 - E_0$ & $\Delta E$ & Length \\ [0.5ex] 
\hline\hline
$0$ & $0.4123$ & $0.4154$ & $200$\\ 
\hline
$0.3$ & $0.7132$ & $0.7132$ & $120$\\
\hline
$0.5$ & $0.7176$ & $0.7007$ &$200$\\
\hline
$0.8$ & $0.2212$ & $0.2147$ & $200$\\
\hline
\end{tabular}
\label{Table1_benchmarks}
\end{table}

\begin{figure}
\centering
\includegraphics[height = 8.75cm, width = 8.75cm, keepaspectratio]{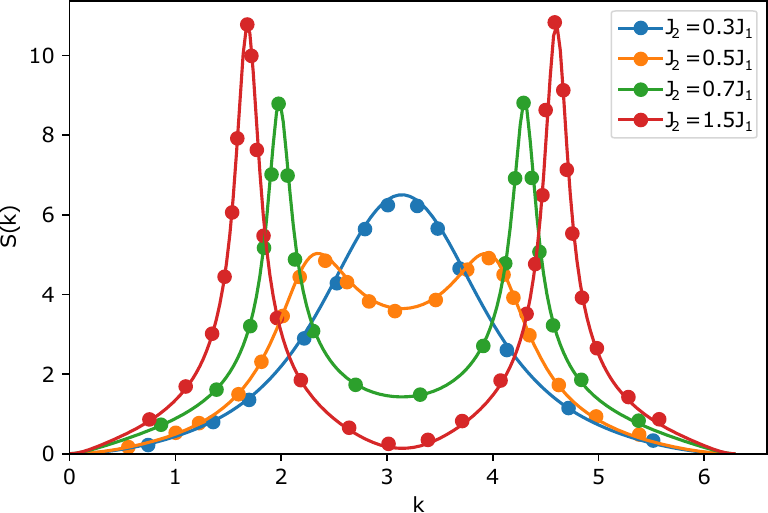}
\caption{Equal-time structure factors ($S^{zz}(k)$) for different $J_2$ interaction strengths. We show a comparison of data obtained from two different approaches - (a)  from Fourier transform of spin-spin correlation functions in the groundstate simulations taken from Ref. \cite{kolezhuk1997variational} (denoted by dots), (b) and  from integrating the DSF $S^{zz}(k,\omega)$ over $\omega$ (plotted as solid lines).}
\label{Fig2Sq}
\end{figure}

Another benchmark for the numerically calculated DSF is to obtain the static structure factor by integrating the DSF in $\omega$ i.e.
 \begin{eqnarray}
 S^{z,z}(k)=\int S^{z,z}(k,\omega)d\omega
 \label{Eq_Sq_factor}
 \end{eqnarray}
 This quantity can be directly calculated from the groundstate of the system by computing equal-time spin-spin correlations and taking a Fourier transform in space. We compared the structure factor obtained from our numerical DSF with that of Ref.\cite{kolezhuk1997variational} in Fig. \ref{Fig2Sq}. There is a good overlap between the two, indicating that the time-evolution in our simulations has remained robust.

\subsection{Single-Mode-Approximation}
\label{SMAsection}

The single mode approximation is used to find the dispersion of a localized excitation of the system \cite{bijl1940lowest, feynman1954atomic, girvin1986magneto}.  The action of a single spin operator on the groundstate describes an excited state, and the energy expectation value of such a state gives the first magnetic moment of the system. Thus, the dispersion of a magnon mode is readily found by taking the expectation value of the energy of an excited state generated from the action of a single spin operator with well-defined momentum on the groundstate of the system.  It is a useful tool to analyse the low-lying excitations of a Hamiltonian. An early successful use of the SMA in spin models was to describe the magnon dispersion in the AKLT model \cite{arovas1988extended, auerbach1998interacting}.  Since the ground state obtained from the DMRG algorithm is very accurate and can be concisely described as an MPS, one can simply extend the SMA formalism to the MPS and obtain the dispersion relation of the magnon modes. 

A single mode with momentum $k$ in a chain of length $N$ is given by:
\begin{eqnarray}
\ket{k} = \frac{1}{\sqrt{N}}\sum_j e^{i k r_j}|\Omega_j\rangle,
\label{Eq_sma_state}
\end{eqnarray}
where the state $|\Omega_j\rangle$ is an excitation on the ground state $|\psi_0\rangle$ at site $j$. The dispersion is readily obtained by:
 
\begin{eqnarray}
\omega (k)= \frac{\bra{k}H\ket{k}}{\bra{k}\ket{k}}-E_0,
\label{Eq_sma_defn}
\end{eqnarray}
where $E_0$ is the ground state energy of the system. Substituting Eq. \ref{Eq_sma_state} into Eq. \ref{Eq_sma_defn}, and expanding the exponentials in the resulting equation into trigonometric functions, one obtains the following relation:
\begin{eqnarray}
\omega(k)=\frac{a_0+\sum\limits_{n=1}^{N-1}a_n\cos(nk)}{1+\sum\limits_{n=1}^{N-1}b_n\cos(nk)} - E_0,
\label{Eq_sma_defn_trig}
\end{eqnarray}
where $b_0 = \sum\limits_j \langle \Omega_j | \Omega_j \rangle$, $a_0 = \sum\limits_j \langle \Omega_j | H | \Omega_j \rangle/b_0 $, $a_n = 2 \sum\limits_j  \langle \Omega_j | H | \Omega_{j+n} \rangle/b_0 $, and $b_n = 2 \sum\limits_j  \langle \Omega_j |\Omega_{j+n} \rangle/b_0 $.
			
In the $J_1 - J_2$ spin-1 chain, there are two possibilities to obtain a single mode dispersion from an excitation on the groundstate:
\begin{figure}
\centering
\includegraphics[width= 9cm, height = 9cm, keepaspectratio]{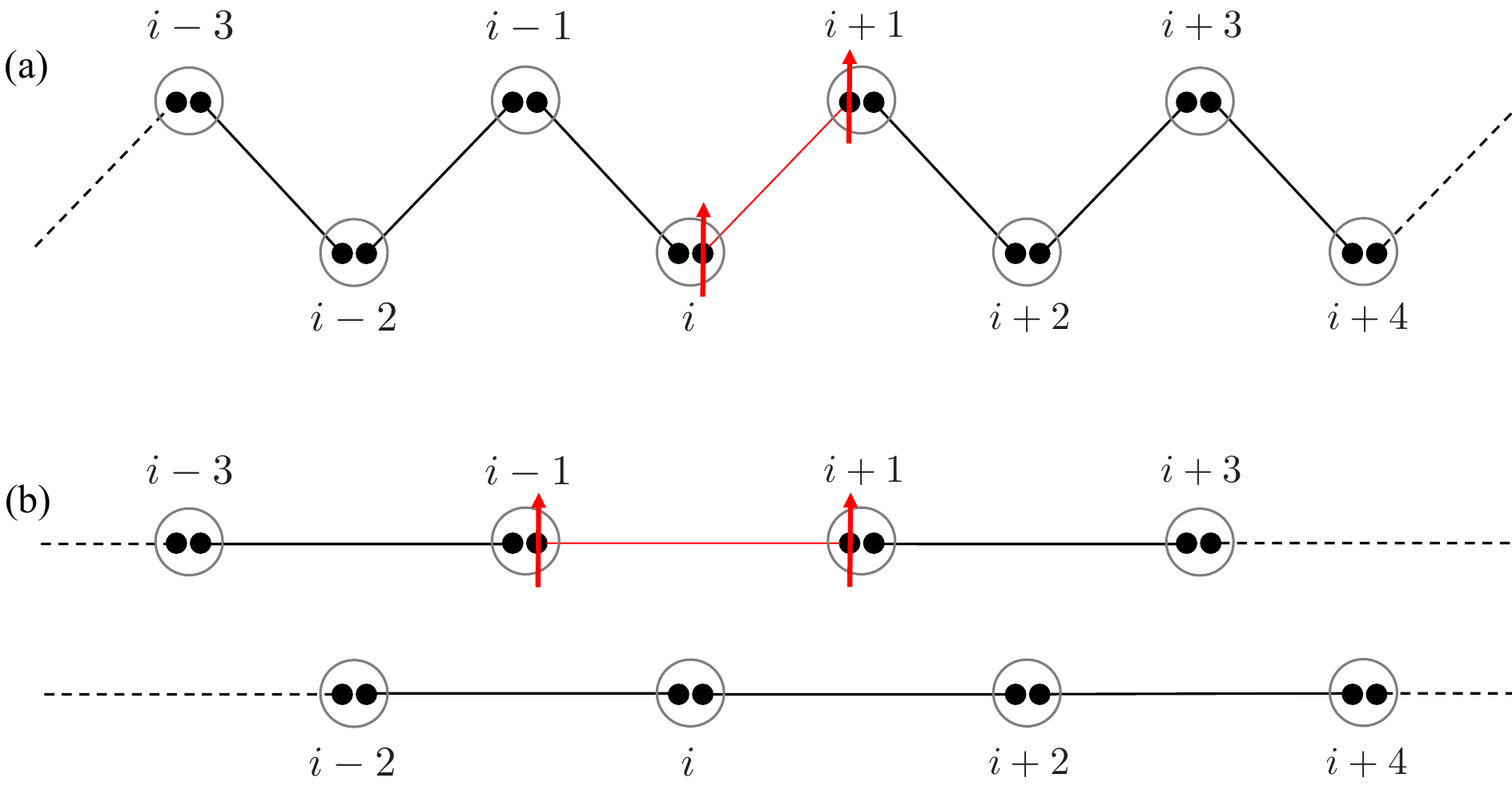}
\caption{An illustration of a single bond-flip of singlet (black bond) to triplet (red bond) in (a) AKLT state, or (b) NNN AKLT state where a nearest-neighbour or a next-nearest-neighbour bond is flipped, respectively. The black dots represent spin-1/2 degrees of freedom while the grey circle enclosing the two spin-$1/2$s at each site represents the physical spin-$1$ degrees of freedom. The bond-flip traveling in the chain with a definite momentum constitutes the magnon excitation in such groundstates.}
\label{Fig3_magnons}
\end{figure}

\begin{enumerate}
\item {\it A local on-site or on-bond excitation} : One can generate a well-defined momentum state by applying $S^z_{-k}$ on the ground state. The ground state of the spin-1 Hamiltonian can be best pictured as an AKLT or NNN AKLT type VBS state when $J_2$ is small or large, respectively. It is readily verified  that the magnon dispersion is obtained by exciting one of the singlet bonds to a triplet bond (see Fig.\ref{Fig3_magnons}). The traveling bond-flip state with a well-defined momentum describes the magnon mode. The application of $S^z_{-k}$ on the VBS state generates a linear combination of the triplet bonds \cite{kolezhuk1997variational, fath1993solitonic}. Therefore, the single spin operator with a given momentum can also determine the magnon dispersion. We have used the groundstate obtained from DMRG and acted the $S^z_{-k}$ operator to describe an approximate magnon state. The energy of such a mode is computed directly by taking the expectation value of the Hamiltonian.

\item  {\it A local domain-wall excitation} : Consider the real space configuration of a wavefunction such that one part is described by the AKLT state while the other is described by the NNN AKLT state. The domain wall, separating these two parts, is an excitation which can disperse in the chain. In order to generate such an excitation, a non-local operator needs to be applied to the state. Determining the expression for such an operator is difficult, so we directly generate these states. The single domain wall excitation is prohibitively expensive in energy if $J_2$ is deep in the Haldane or  NNN Haldane phase as it would require part of the wavefunction given by the other phase's groundstate but it is the lowest excited state at the phase transition point in the spin-1 chain since the two types of VBS states have equal energies at that point (see Fig.\ref{Fig7_spinon_fitting} b).  If two such domain walls are placed adjacent to each other, one recovers the triplet-bond picture describing a magnon in the previous scenario.  Therefore, these domain walls are fractionalized quasi-particles and will be referred to as spinons because they carry half the spin of the magnon, quite similar to the domain walls in spin-1/2 chains being referred to as spinons. The SMA study of the spinons will turn out to be instructive in understanding the spectrum close to the phase transition.
\end{enumerate}
	
\section{Dynamical Structure Factor of the $J_1-J_2$ spin-1 chain}
\label{numerical_spectral_functions_J1_J2}
			
\begin{figure*}[!t]
\centering
\includegraphics[width=18cm, height = 18cm, keepaspectratio]{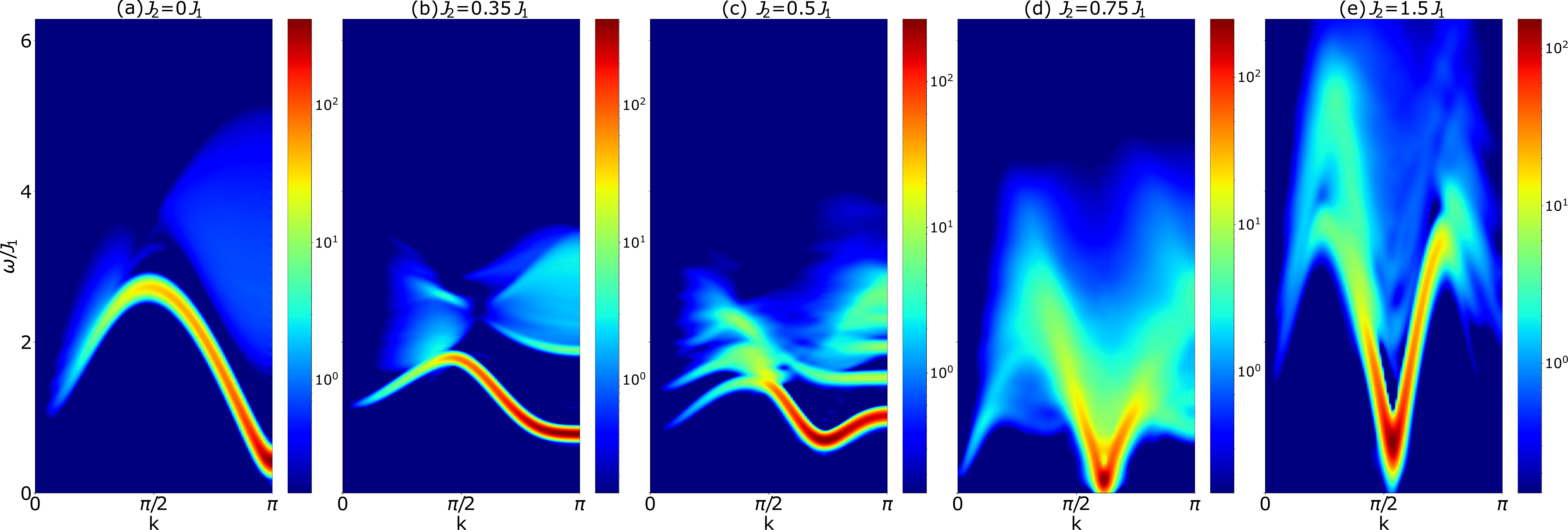}
\caption{(a-e) Dynamical Structure Factor ($S^{zz}(k,\omega)$) as $J_2$ is varied. In logarithmic colour scale, one observes the single magnon mode and two- and three-magnon continua when $J_2 = 0$. As $J_2$ is increased, the magnon and its associated continua smoothly evolve into deconfined spinons at $J_2\approx 0.76 J_1$. Beyond the first-order transition, in the extreme large limit of $J_2$, one again finds a prominent magnon mode with halved period than before because the unit cell size is doubled for a $J_2$ chain. }
\label{Fig4_spectra}
\end{figure*}

Using the time evolution algorithm, we calculated the DSF $S^{zz}(k,\omega)$  of the $J_1-J_2$ chain for different values of $J_2$ ranging from $0$ to $1.5J_1$. We discuss various features of the spectra by dividing the results into three parts : (a) deep in the Haldane phase, where $J_2$ ranges from $J_2 = 0$ to $J_2 = 0.6J_1$; (b) vicinity of the phase transition, where $J_2$ ranges from $J_2 = 0.7J_1$ to $J_2 = 0.8J_1$; (c) deep in the NNN Haldane phase, where $J_2>J_1$ . An overview of the spectra can be found in Fig. \ref{Fig4_spectra}. As we shall see, many features can be interpreted as magnons, spinons, or composite particles. 

\subsection{ $\mathbf{0\leq J_2/J_1\leq 0.6}$}
\label{multibd}
In the Haldane phase, the ground state is pictured as a VBS-type state.  Therefore, flipping a single bond between adjacent sites generates a subspace of single-bond flipped states. Diagonalizing the Hamiltonian in this subspace in momentum basis results in a single magnon mode, which is the lowest-lying excitation. Using the dispersion of the magnon one can create the continua of multiple magnons. 

\begin{figure*}
\centering
\includegraphics[width=18cm , height = 18cm, keepaspectratio]{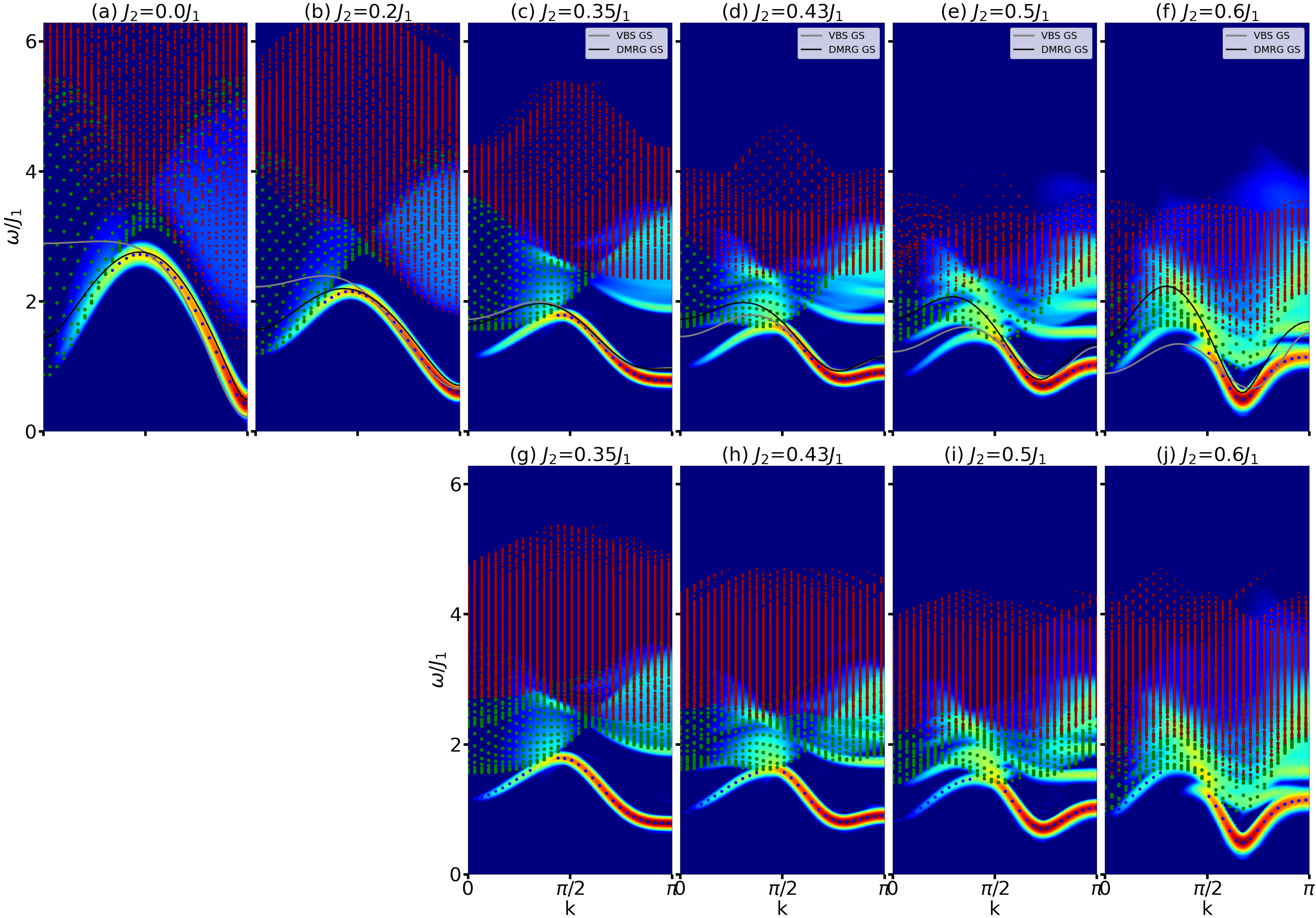}
\caption{ Multi-magnon continua on top of DSF. The  single magnon mode is shown with dark-blue dots. The two-magnon continuum obtained by taking two individual single magnons is shown in green. Similarly a three-magnon continuum is created from the single magnon mode and is shown in red. The multi-magnon continua can be generated from the single magnon (described in the main text). For $J_2<0.5 J_1$, one finds good agreement between the SMA dispersion computed on the groundstate obtained from DMRG, and the magnon dispersion obtained from the AKLT state at low energies. This indicates that the AKLT state is a good approximation to the groundstate of the spin-1 chain for small values of $J_2$.}
\label{Fig5_bs}
\end{figure*}
		
\begin{figure}

\centering

\includegraphics[width = 8.75cm, height=8.75cm, keepaspectratio]{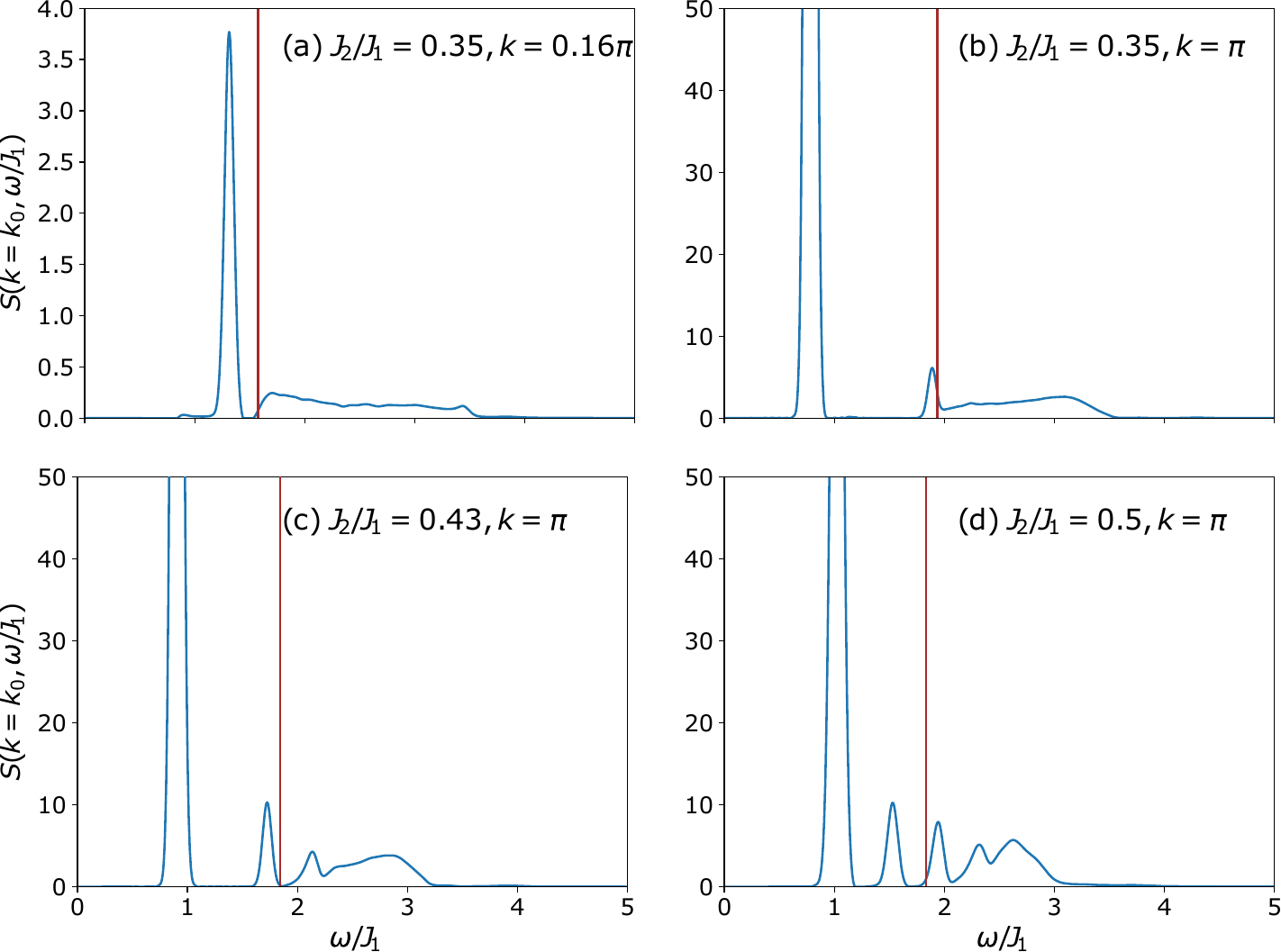}
\caption{ Section cuts of DSF for selected $J_2$ values away from the phase transition point at selected $k$ values to illustrate the bound state of two magnons lying outside of the two-magnon continua. The lower extent of the continua for the particular $k$- value is indicated by a vertical line at the $\omega$ value. The excitations below this line are a well-defined mode.}
\label{Fig6_section_cuts}
\end{figure}

At $J_2 =0$, one finds a well-defined gapped magnon mode. The dispersion of the mode is also correctly described by SMA; however, at $k \approx 0.33\pi$, the mode enters the two-magnon continuum \cite{white2008spectral}. Therefore, although the SMA gives a description of the mode beyond the entry point into the continuum, this cannot be trusted: The excitation exists as a well-defined mode for a range of wave-vectors $k$, and it gives rise to the non-interacting continuum of two or three magnons. In this way, one has a consistent description of the full spectra found in the DSF. 

For $J_2>0$, one finds that the magnon mode is separated from its multi-magnon continua for some range of momentum ($k$) values. Similar to the $J_2 = 0$ case, the magnon enters the continua at some $k_0$ value. This is marked by noticeable loss in the spectral intensity of the mode, however one can continue to follow the mode to small $k$ values. The interaction of the two-magnon continuum with the single magnon mode results in a bound state of two magnons which appears as a resonance in the two-magnon continuum for $0<J_2/J_1<0.35$ and as a separate mode below the two-magnon continuum for $0.35\leq J_2/J_1 \leq 0.6$ at small $k$ values ($0\leq k\leq k_0$). In order to clarify this point,  we tracked the magnon mode in the section cut at $k \approx 0.16\pi$ of the DSF for $J_2 = 0.35 J_1$ (see Fig.\ref{Fig6_section_cuts}(a)). It demonstrates that a single mode is separated from the lower bound of the continuum, with a line-width that is only limited by the frequency resolution of the DSF. 

This adds a subtlety to the interpretation of the non-trivial higher energy excitations which lie outside the continua because they can be seen as the bound states of single magnons or the bound states of the single lowest energy mode comprising of a single magnon in $k_0<k<2\pi-k_0$ and the bound state mode in $0\leq k\leq k_0$. We identify the different origins of the bound states by sequentially following two steps : (i) identifying which modes lie outside the non-interacting two-magnon continuum, and (ii) identifying which among these modes lie outside the continuum generated by the single lowest energy mode seen in the DSF. 
 
\begin{enumerate}
\item {\it Modes lying outside of multi-magnon continua} : In Fig.\ref{Fig5_bs} (a-f), we plot the non-interacting $n$- magnon continua on top of the computed DSF (n = 2 is shown in green and n = 3 is shown in red). It covers the full energy range of the observable spectral weights in the DSF.  The modes which lie outside of the multi-magnon continua are the bound states of magnons. In general,  a $n$-magnon boundstate at wave-vector $K = \left(\sum_{j}k_{j}\right)\mathrm{mod}~2\pi$ has an energy $\sum_{j =1}^{n}\omega_j(k_j) +E_b$, where $E_b$ is the binding energy and is negative. We clearly demonstrate that the mode $0<k<k_0$ is a bound state of two magnons and its binding energy increases with increasing $J_2$. At $J_2 = 0.6J_1$, one finds a second bound state separated from the continuum in this momentum range. A similar observation can be made at $k = \pi$, where an increasing number of bound states of magnons separate out from the three-magnon continuum as $J_2$ increases.

 \item {\it Modes lying outside of continua generated from the lowest energy mode} : The SMA tracks the dispersion of the magnon and, therefore, is unable to track the lowest-energy mode when its nature changes in the range $0\leq k \leq k_0$. We track the mode by following the maxima of the spectral peak in the DSF at the lowest energy and construct the multi-mode continua out of it.  In  Fig. \ref{Fig5_bs}(g-j),  the continua are plotted on top of the DSF and as $J_2$ is varied in the range $0.35\leq J_2/J_1\leq 0.6$, one sees a clear separation of the second lowest energy mode at $k=\pi$ from the multi-mode continua. It is interpreted as a three-magnon bound state.  We track the mode for different $J_2$ values by plotting the section cuts of their corresponding DSF at $k=\pi$ (see Fig. \ref{Fig6_section_cuts} b, c, d). The spectral peak corresponding to the lowest energy bound state mode is lower in energy than the lower bound of the multi-magnon continua. Since the other apparent higher-energy bound states of magnons lie within the multi-mode continua, these can be interpreted to be resonances in the DSF. In the same spirit, for small values of $k$ i.e. in the range of momenta  $0<k<k_0$, the DSF plots for $J_2 = 0.5J_1,~ 0.6 J_1$ ( Fig. \ref{Fig5_bs} i,j) show another three-magnon bound state mode which pops out of  the continua. Both bound states gain in binding energy as $J_2$ increases.
Note that for $J_2=0.6J_1$, the lowest bound state for small $k$ has lost so much intensity that it is not possible any more to follow it down to $k=0$. As a consequence, we could not include it when plotting the multi-mode continuum, resulting in a smaller continuum close to $k=\pi$ (see Fig.\ref{Fig5_bs}j) than anticipated from the continua at smaller values of $J_2$ (see Fig.\ref{Fig5_bs}h,i). But this is an artefact because the mode still exists, and the intensity close to $k=\pi$ above the main magnon branch and the lowest bound state (which nearly merges with the main magnon branch at $k=\pi$) must be part of a continuum, as for $J_2/J_1=0.43$ and $0.5$.
  
\end{enumerate} 
		
\subsection{Vicinity of the phase transition ($\mathbf{0.7\leq J_2/J_1\leq 0.8}$)}
\label{neardomain}

\begin{figure*}

\centering
\includegraphics[width=18cm, height = 18cm, keepaspectratio]{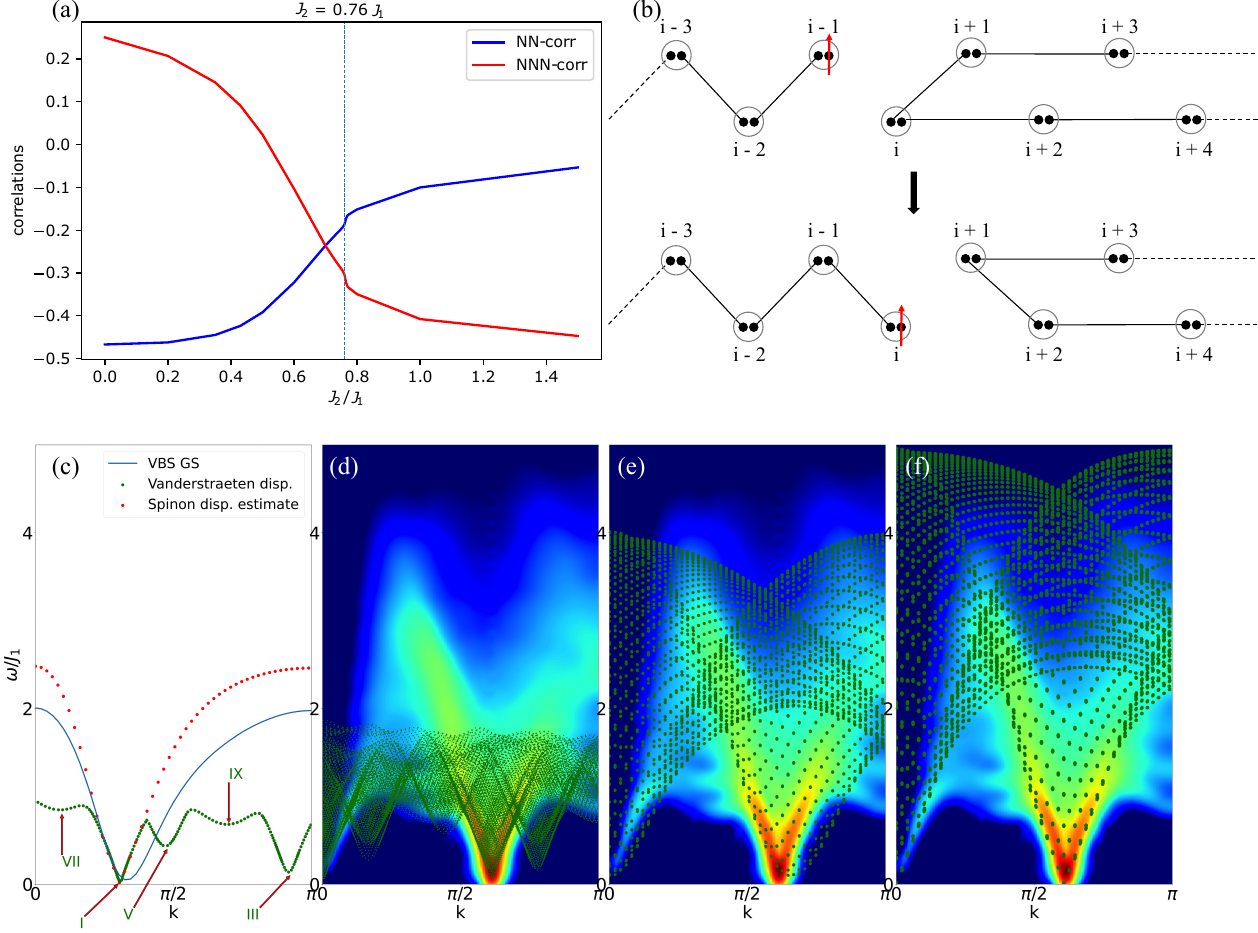}
\caption{(a) Determination of the $J_2$ interaction strength at which the first-order phase transition by locating the drop/jump in the nearest-neighbour ($\langle S^z_iS^z_{i+1} \rangle$) and next-nearest-neighbour ($\langle S^z_iS^z_{i+2} \rangle$)  correlations. (b) A cartoon description of the domain wall separating the two competing groundstates - AKLT and NNN AKLT state. The spin-$1/2$ degree of freedom (shown in red) is indicative but in practice it is $x$-polarized. The choice of the polarization of the unparied spin-$1/2$ does not affect the dispersion of the domain wall. The black arrow indicates the hopping of the domain wall by a single lattice site. The dispersion of the domain wall (spinon) is calculated with the SMA approach. (c) Fitting of the SMA dispersion relation of single spinon with the dispersion determined by quasi-particle ansatz at the first-order phase transition point, such that (i) the location of the minima of the dispersion agrees with the quasi-particle ansatz estimate of the dispersion, (ii) the spinon dispersion agrees well with the low energy spinon dispersion obtained with the quasi-particle ansatz, and (iii) the amplitude of the dispersion is such that the resulting spinon continuum spans the full extent of the spectral function. The extra features (labelled with roman numerals) in the quasi-particle ansatz determined dispersion are artefacts because it tracks total spin-1/2 and it belongs to minimas of $(2n+1)$-spinon continua.  (d) The two-spinon continuum generated from the quasi-particle ansatz single spinon dispersion put on top of the DSF evaluated using tDMRG at $J_2 = 0.76J_1$. It does not match so well indicating a much simpler dispersion of the spinon. (e) The two-spinon continua generated from spinon-dispersion obtained from the SMA analysis of the domain wall between AKLT and NNN AKLT state. The groundstate at the phase transition is more complex but the simple form of dispersion already captures the broad features of the numerically determined DSF, (f) The two-spinon continuum generated from the simple spinon dispersion covers the full spectral function with the minima matching the one seen in the DSF. }
\label{Fig7_spinon_fitting}
 \end{figure*}

The $J_1-J_2$ spin-1 chain undergoes a first-order transition \cite{kolezhuk1996first}. We compute the nearest-neighbour (NN) and next-nearest-neighbour (NNN) correlation functions with varying $J_2$ and find that at the phase transition point there is a simultaneous  drop and jump respectively (see Fig. \ref{Fig7_spinon_fitting} a). This is due to a change in the type of groundstate with $J_2$ interaction strength - the groundstate in the Haldane phase is an AKLT type VBS order and the groundstate in the NNN Haldane phase is a NNN AKLT type VBS order. Thus, by computing  the NN and NNN correlation functions, one locates the first-order phase transition to be at $J_2 = 0.76J_1$. It has been demonstrated by L. Vanderstraeten {\it et.al.}  \cite{vanderstraeten2020spinon} that the single-domain wall excitation is the lowest energy excitation and the gap does not close at the phase transition point. The spinon dispersion relation was obtained at the phase transition point by using the quasi-particle ansatz formulation of  MPS (see Fig.\ref{Fig7_spinon_fitting} c, d). One can generate the two-spinon continuum using this spinon dispersion, but it does not agree with the tDMRG generated DSF  : 
\begin{enumerate}
\item Although the lowest minima peak of the continua aligns with the DSF minima, the two-spinon continua has many more local minima peaks than the DSF spectra. 
\item  The continua covers only up to $~2J_1$ energies while the spectral weight in the DSF ranges up to $~4J_1$.  Since the spinons are deconfined at the phase transition point, one would have expected them to span the full energy range of the DSF spectra.
\end{enumerate}
\par
Moreover, if we denote  by $k_{\mathrm{min}}$ the wave-vector of the global minimum of the spinon dispersion, then one finds that the other local minima of spinon dispersion are located at $(2n+1)k_{\mathrm{min}}$ where, $n$ is an integer.  Similarly, the local minima energies lie above $\omega\approx (2n+1)\Delta$, where $\Delta$ is the spinon energy gap. Now, since the quasi-particle ansatz picks the lowest lying energy state for each momentum value in the spin-1/2 sector of the excitation spectrum, it captures the minima of $(2n+1)$ spinon continua which lie in the total spin-1/2 sector (see Fig. \ref{Fig7_spinon_fitting} c). So, the curve for the spinon dispersion extracted from Ref.  \cite{vanderstraeten2020spinon} only represents the true spinon dispersion in the vicinity of the global minimum.

\subsubsection{SMA of the domain wall}
\label{VBS_domain_wall_dispersion}

Since the two competing ground states at the phase transition are approximately seen as the VBS and the NNN VBS state, one can attempt a SMA calculation on a wave function where one part is the VBS state and the other part is the NNN VBS state. Since these states can be expressed as MPS exactly, we use the SMA approach to calculate the dispersion relation. We provide the details of representing a domain wall state as an MPS in Appendix \ref{spinoncalc}. The dispersion obtained is plotted as a blue line in Fig.\ref{Fig7_spinon_fitting}c. The continua generated from such a dispersion is then compared with the DSF spectra in Fig. \ref{Fig7_spinon_fitting}e. The position of the local minima of the continuum is at $k\sim 0.6580\pi$ as compared to the location of the DSF minima at $k = 0.6129\pi$.  Furthermore, the spinon continuum extends up to much larger energies of the order of $4J_1$, in better agreement with the full DSF.  Although it improves on the quasi-particle ansatz in obtaining a qualitatively more reasonable dispersion, it does not have quantitative accuracy as the DSF.  This is because the system in the vicinity of the phase transition point has a longer correlation length than the AKLT or NNN AKLT states. Numerically, the real ground states will be described by an MPS of larger bond dimension than the AKLT states we considered in the calculations. Still, such an analysis highlights that the spinon dispersion is $\pi$-periodic and is of much simpler form when compared against the dispersion obtained from quasi-particle ansatz.\par

\begin{table}
\centering
\caption{Coefficients of Eq. \ref{Eq_sma_defn_trig} for the spinon dispersion ( up to four terms) for the domain wall dispersion on a VBS state denoted by $a_i, ~b_i$, and the fitted spinon dispersion denoted by $\tilde{a}_i, ~\tilde{b}_i$.}
\begin{tabular}{||c c c c c||} 
\hline
$i$ & $a_i$ & $b_i$  & $\tilde{a}_i$ & $\tilde{b}_i$\\ [0.5ex] 
\hline\hline
$0$ & $-199.333$  & - & $-198.14$ &  -\\ 
\hline
$1$ & $-0.333$  & $0$ & $327.011$ &  $-1.651$\\
\hline
$2$ & $66.667$  & $-0.3333$ &  $-184.517$ &  $0.932$\\
\hline
$3$ & $44.611$  & $-0.222$ & $63.682$ &  $-0.322$\\
\hline
$4$ & $7.4444$  & -0.037 & $-11.65$& 0.059\\
\hline
\end{tabular}
\label{Table2_SMA_coeffs}
\end{table}

\subsubsection{Fitting the SMA dispersion of spinon}
The SMA could explain qualitatively the spectral features at the phase transition and points to a simple form of the dispersion relation with only one minimum. Starting from the mathematical form of the SMA dispersion on the VBS state, one can modify the coefficients such that the resulting spinon dispersion accurately covers the low energy features of the continua in the DSF. Since the quasi-particle ansatz  determines the lowest excitation accurately, one can compare the position of the minimum and the slope of the dispersion with respect to the wavevector $k$ with the SMA's analytical form given by
\begin{eqnarray} 
\omega(k)=\frac{\tilde{a}_0+\sum\limits_{n=1}^{n=4}\tilde{a}_n\cos(nk)}{1+\sum\limits_{n=1}^{n=4}\tilde{b}_n\cos(nk)} - E_0.
\label{Eq_fitting_sma_ansatz}
\end{eqnarray}      

A minimum of four cosine terms is needed to obtain a good fit with the low-lying energy dispersion from the quasi-particle ansatz. There is a further constraint on the amplitude of the spinon dispersion such that the lower bound of the resulting continuum lies on the plausible  location of the continua in the DSF. We extracted this amplitude directly from the DSF for $J_2=0.76 J_1$ by generating the two-spinon continuum for a range of amplitudes and comparing its lower bound with the inflection point just below the highest spectral peak at  $k =\pi$ (the method is described in detail in Appendix \ref{spinonheight}). The upper bound of the two-spinon continuum comes out to be just below $\omega \sim 5J_1$ by which $\sim 99.5\%$ of the spectral weight of the DSF has been exhausted. Thus, with this determination of the spinon dispersion, the continuum spans almost the full range of the DSF.
The plot of the resulting spinon dispersion is compared with other determinations discussed earlier in Fig. \ref{Fig7_spinon_fitting}(c). The two-spinon continuum generated from the newly fitted spinon dispersion compares very well with the DSF for $J_2=0.76J_1$(see Fig. \ref{Fig7_spinon_fitting}f).\par

\subsubsection{Comparison between spinon dispersion and tight-binding model }
\label{compare_VBS_domain_wall_fitted_spinon}

In order to get an intuitive understanding of the dispersion of the spinon, let us try to express it as the dispersion of a tight-binding model.  The form of the SMA dispersion consists of a numerator and a denominator which can be non-trivial depending on the Hamiltonian of the system. 
The spinon state is constructed by leaving a spin-$1/2$ unpaired at the $j$-th site which separates the AKLT and NNN AKLT state (see Fig. \ref{Fig7_spinon_fitting}b). Such a state is denoted by $|\Omega_j\rangle$. The denominator in the SMA dispersion relation is the squared norm of the travelling spinon state (denoted by $|\Omega_{-k}\rangle$) and can be expanded as:
\begin{eqnarray}
\mathcal{O}_{k} &=& \langle \Omega_{k}|\Omega_{-k}\rangle = \sum_{i,j}e^{-ik(r_i-r_j)}\langle \Omega_j|\Omega_i\rangle \nonumber\\
&=& \sum_{i,j}e^{-ik(r_i-r_j)}\begin{bmatrix}\mathcal{O}\end{bmatrix}_{ij}
\label{Eq_overlap_mat}
\end{eqnarray}
where,  $\begin{bmatrix}\mathcal{O}\end{bmatrix}$ is the overlap matrix .

The overlap matrix can be non-trivial. The energy values one finds can be seen as the eigenvalues of an effective Hamiltonian ($\tilde{H}$) defined in the subspace spanned by non-orthogonal basis $\lbrace |\Omega_i\rangle ,~ 0\leq i\leq N\rbrace$. Consider an eigenstate $|\psi\rangle$ of the effective Hamiltonian, then it can be written as a superposition of the non-orthogonal basis as follows: 
\begin{eqnarray}
|\psi\rangle =  \sum_{i}\psi_i|\Omega_i\rangle
\label{Eq_eff_eigenstate}
\end{eqnarray}
This wavefunction is also an eigenstate of the full Hamiltonian (i.e. $H|\psi\rangle = \omega|\psi\rangle$). Taking overlap with $\langle\Omega_j|$ on both sides one finds  : 
\begin{eqnarray}
&&\sum_{i}\langle \Omega_j| H |\Omega_i\rangle \psi_i = \omega\sum_{i}\langle \Omega_j|\Omega_i\rangle\psi_i\nonumber\\
 &&\sum_k\left[\mathcal{O}^{-1}\right]_{ik}\langle\Omega_k|H|\Omega_j\rangle\psi_i = \omega\psi_i,
 \label{Eq_eff_H_derivation}
\end{eqnarray}
where one takes the inverse of the overlap matrix to return to an eigenvalue problem for the effective Hamiltonian $\tilde{H}$. The form of the effective Hamiltonian can be read off from the last line to be:

\begin{eqnarray}
\tilde{H}= \sum_k\left[\mathcal{O}^{-1}\right]_{ik}\langle\Omega_k|H|\Omega_j\rangle
\label{Eq_eff_H}
\end{eqnarray}

  It can be diagonalized in the momentum basis and the energies are given by 
\begin{eqnarray}
\omega(k)&=&\bra{\Omega_{N/2}}H\ket{\Omega_{N/2}}\\
& + & \sum_{i=1,i\neq N/2}^{i=N} \bra{\Omega_{N/2}}H\ket{\Omega_{i}} e^{i k(N/2 - i)}\nonumber
\end{eqnarray}
Since the coefficients for $i\neq N/2$ satisfy $\bra{\Omega_{N/2}}H\ket{\Omega_{N/2+i}} = \bra{\Omega_{N/2}}H\ket{\Omega_{N/2 - i}}$, this can be rewritten as 
\begin{eqnarray}
\omega(k) &=&\gamma_0 + \sum_{\delta=1}^{N/2 -1 }2\gamma_\delta \cos(k\delta)
\label{Eq_eff_H_eigenvalues}
\end{eqnarray}
where $\gamma_0=\bra{\Omega_{N/2}}H\ket{\Omega_{N/2}}$ and $\gamma_\delta=\bra{\Omega_{N/2}}H\ket{\Omega_{N/2+\delta}}$.\par

For describing the motion of the spinon on the ground state, we generated a subspace by placing the domain wall state separating the AKLT state and NNN AKLT state at different bonds of the chain (see Fig. \ref{Fig7_spinon_fitting}b). We can expand the Eq. \ref{Eq_sma_defn_trig} into sum of  cosines and read off the coefficients $\gamma_\delta$. These can be found in Table \ref{Table3_hopping_coeffs}.
From the discussion about obtaining a well-fitted spinon dispersion, we had determined the coefficients $\tilde{a}$ and $\tilde{b}$ in Eq. \ref{Eq_fitting_sma_ansatz}. They are tabulated in Table \ref{Table2_SMA_coeffs}. We carry out the numerical expansion of such a dispersion into sum of cosines and tabulate the coefficients $\tilde{\gamma}_{\delta}$ in Table \ref{Table3_hopping_coeffs}

\begin{table}
\centering
\caption{Comparison of $\gamma_\delta$ - tight-binding hoppings generated from the dispersion of the domain wall between the AKLT and NNN AKLT state and $\tilde{\gamma}_\delta$ - the tight-binding hoppings generated from the fitted spinon dispersion.}
\begin{tabular}{||c c c||} 
\hline
$\delta$ & $\gamma_\delta$ & $\tilde{\gamma}_\delta$ \\ [0.5ex] 
\hline\hline
0 & 1.333 & 1.716\\ 
\hline
1 & -0.222 & -0.293\\
\hline
2 & 0.306 &0.273\\
\hline
3 & 0.259 & 0.32\\
\hline
4 & 0.099 & 0.16\\
\hline
\end{tabular}
\label{Table3_hopping_coeffs}
\end{table}

\begin{figure}
\centering
\includegraphics[width=8.5cm, height = 8.5cm, keepaspectratio]{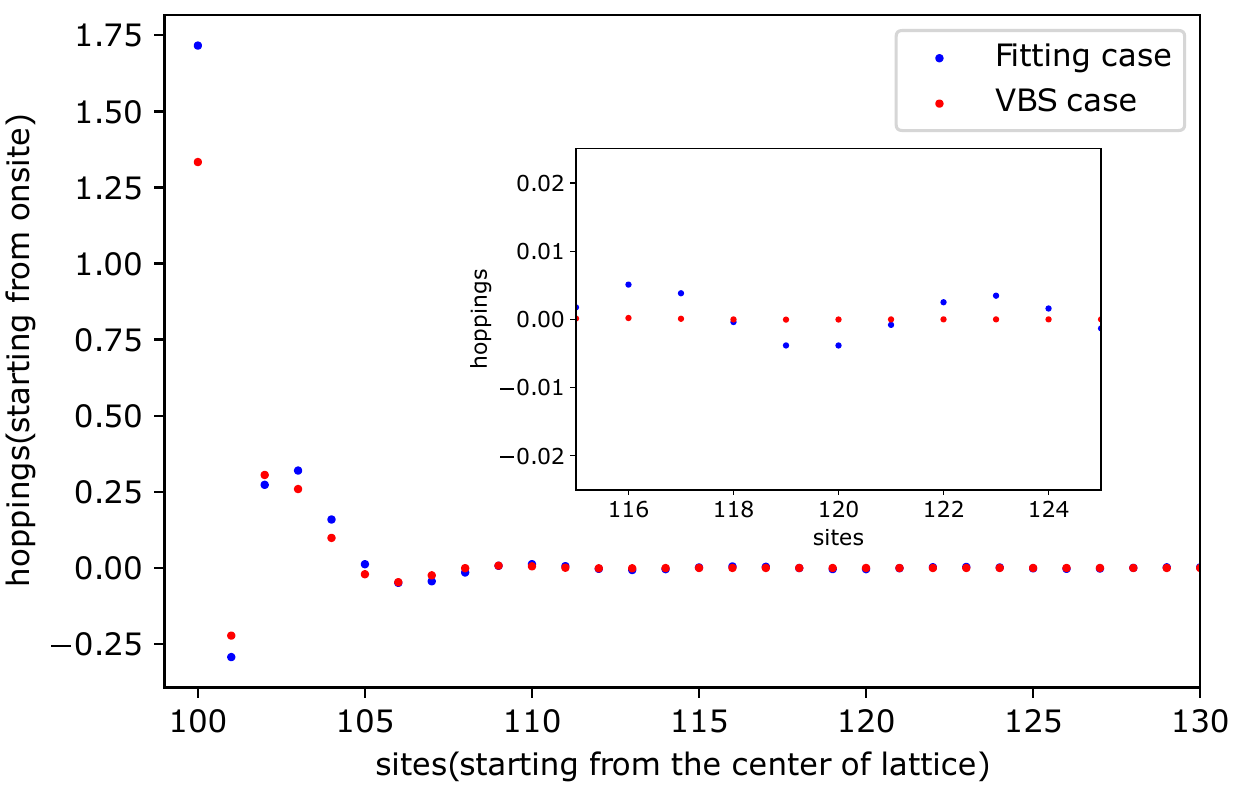}
\caption{ Comparison of the tight-binding hopping coefficients $\gamma_\delta$ corresponding to the domain-wall between the AKLT and NNN AKLT state with the tight-binding hopping coefficients $\tilde{\gamma}_\delta$'s corresponding to the fitted spinon dispersion for $J_1 - J_2$ spin-1 chain at the phase transition. The simplified picture of the domain-wall at the phase transition generates a set of tight-binding coefficients which decay to zero but the ones determined from the fitted spinon dispersion relation do not decay as quickly. In the inset, we show a zoomed in picture and show that the effective long-range hoppings are non-zero for the fitted spinon dispersion.   This is because of the sharp slope of the dispersion around the minima (see Fig. \ref{Fig7_spinon_fitting}c): one needs more long-range hopping terms to realize such a slope.} 
\label{Fig8_hopping}
\end{figure} 

The form of the dispersion relation of Eq. \ref{Eq_eff_H_eigenvalues}  with coefficients $\gamma_\delta$ or $\tilde{\gamma}_{\delta}$ is similar to the dispersion obtained from a tight-binding model with further neighbour hoppings. From both spinon dispersions, one finds that the effective tight-binding model has a negative first-neighbour hopping term and a positive second-neighbour hopping term (see Fig. \ref{Fig8_hopping}). It is the competition between these leading hopping terms that leads to the minimum of the dispersion at an incommensurate wave vector.

\subsubsection{Spectra near the transition point}
\label{neardomain}

\begin{figure*}

\centering
\includegraphics[width=18cm, height = 18cm, keepaspectratio ]{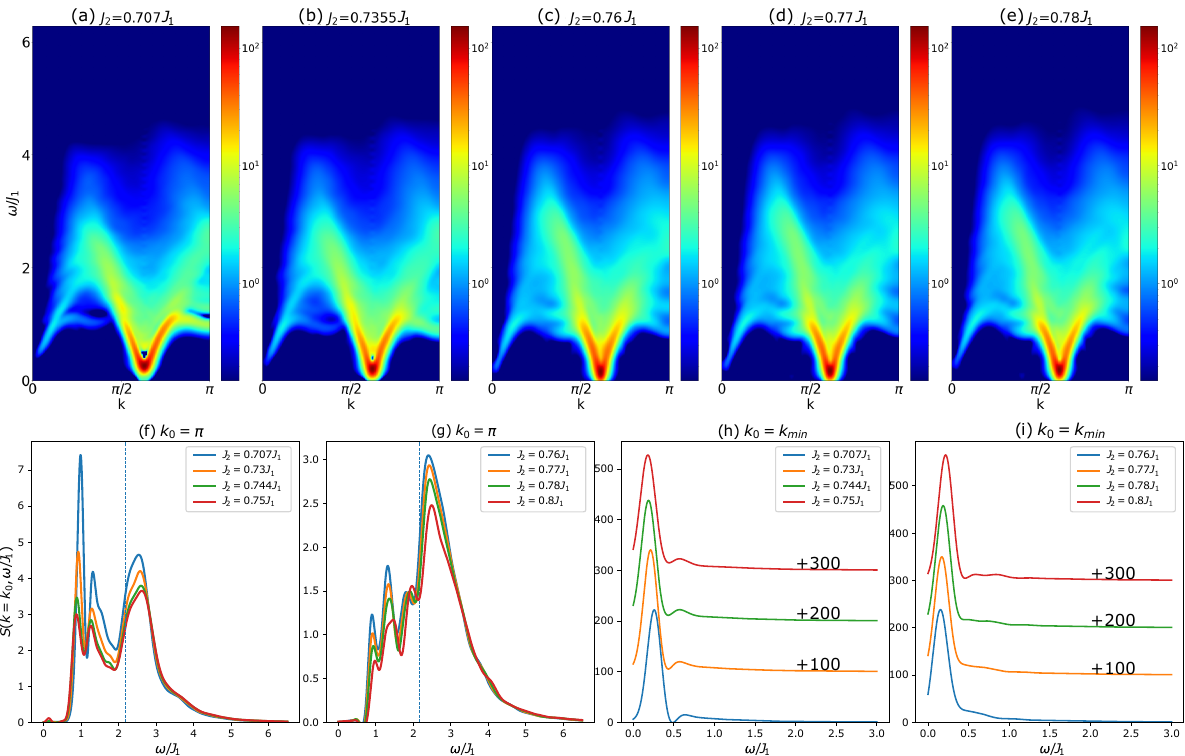}

\caption{(a - e) DSF of the $J_1 - J_2 $ spin-1 chain about the phase transition point.  Section cuts of the DSF with (f) $J_2 \leq 0.75J_1$, and (g) $J_2\geq 0.76J_1$ at $k = \pi$. The lower energy boundary of the two-spinon continuum at $J_2 = 0.76J_1$ is indicated by dashed line.   One notes that the number of separate modes lying outside of the continua increases before and after the transition. These are magnon excitations on the domains constituting the groundstate at the phase transitions.  Similarly, we show the section-cuts of the DSF with (h) $J_2\leq 0.75 J_1$, and (i) $J_2\geq 0.76 J_1$ at $k = k_{\mathrm{min}}$, where $k_{\mathrm{min}}$ is the wavevector corresponding to the global minimum of the continua. It shows multiple modes beyond the main peak, which are interpreted to be bound states of the spinons. They arise due to the confinement potential two spinons experience, because slightly away from the transition point one of the competing groundstates has lower energy-density than the other. Therefore, spreading the domain walls costs energy and thus the continua then proliferates into bound states of the spinons. However, the energy levels of these states are so close to each other that it becomes very difficult to resolve the two states in the numerical DSF, except for the low-lying bound states of spinons where the difference is larger. }
\label{Fig9_deconfinement}
\end{figure*}

At the first-order phase transition point, the two ground states, the Haldane type and the NNN Haldane type, have the same energy densities. The domain walls between these two possible groundstates are deconfined and this results in a continua as seen in the DSF simulations at $J_2 = 0.76J_1$. The section cut at the $k$ value where the spinon continuum reaches a minimum shows a spectral peak at low energies followed by the decay of the spectral weight into a long tail for larger $\omega$ (see Fig. \ref{Fig9_deconfinement} c, i). 
When the $J_2$ interaction strength is tuned away from the phase transition slightly, a low-energy mode splits off the continuum. This is particularly clear for smaller values of $J_2$  (see Fig. \ref{Fig9_deconfinement} a, b, h), but the trend is also visible for larger $J_2$ (see Fig. \ref{Fig9_deconfinement}  i). This is due to the confinement of the domain walls: away from the transition, one of the ground states becomes energetically more favourable than the other and the two domain walls cannot be separated by arbitrary distances on the chain without paying an energy cost.  This energy cost increases proportionally to the distance between the domain walls and can be visualized as a confining potential (i.e. $V (x)\propto  ((J_2-J_{2, t})/J_1)x$, where $J_{2, t}$ is the phase transition point) for the otherwise free spinons. The competition between the lowering of  kinetic energy due to the propagation of spinons and the energy cost due to the effective confining potential results in two-spinon bound states. Note that we could only unambiguously resolve the lowest one. It could be that with a better resolution it might be possible to identify other bound states. However, since the two-spinon continuum is also expected to have a minimum at $k=0$ in addition to that visible in the DSF at $k=0.6129 \pi$, the higher bound states might overlap with the continuum built by pairs of low-energy bound states with wave-vectors close to $k=0$ and $k=0.6129 \pi$ respectively, turning the higher-energy bound states into resonances.

In addition to the bound states of spinons, the system also exhibits magnons and boundstates of magnons. At the transition point, there are modes that lie outside of the two spinon continuum (see Fig. \ref{Fig7_spinon_fitting}f). As $J_2$ is tuned, these modes smoothly evolve into well-defined modes. Since the domain walls separate out the domains described by the VBS and the NNN VBS-like states, it is possible to locally excite bound states of magnons in one of the domains. The associated spectral features can be tracked for $J_2$ in the vicinity of the transition at $k = \pi$ by plotting the section cuts in Fig.\ref{Fig9_deconfinement} f,g. We can distinguish at least three spectral peaks below the two-spinon continuum lower bound and the spectral intensity of the modes increases as $J_2$ decreases from the NNN Haldane phase to the Haldane phase. The lowest energy spectral peak increases in intensity in the section cut at $k=\pi$ and smoothly evolves into the bound state as seen in the DSF for $J_2 = 0.7 J_1$ (in Fig.\ref{Fig5_bs}g). These bound states can be traced in the DSF for smaller $J_2$ values ($0.35\leq J_2/J_1\leq 0.7$) as the gap between two-particle continuum and the modes reduces to eventually turn into resonances in the spectra. 

 \subsection{$\mathbf{J_2/J_1 > 0.8}$}
\begin{figure}
\centering
\includegraphics[width=9cm, height = 9cm , keepaspectratio]{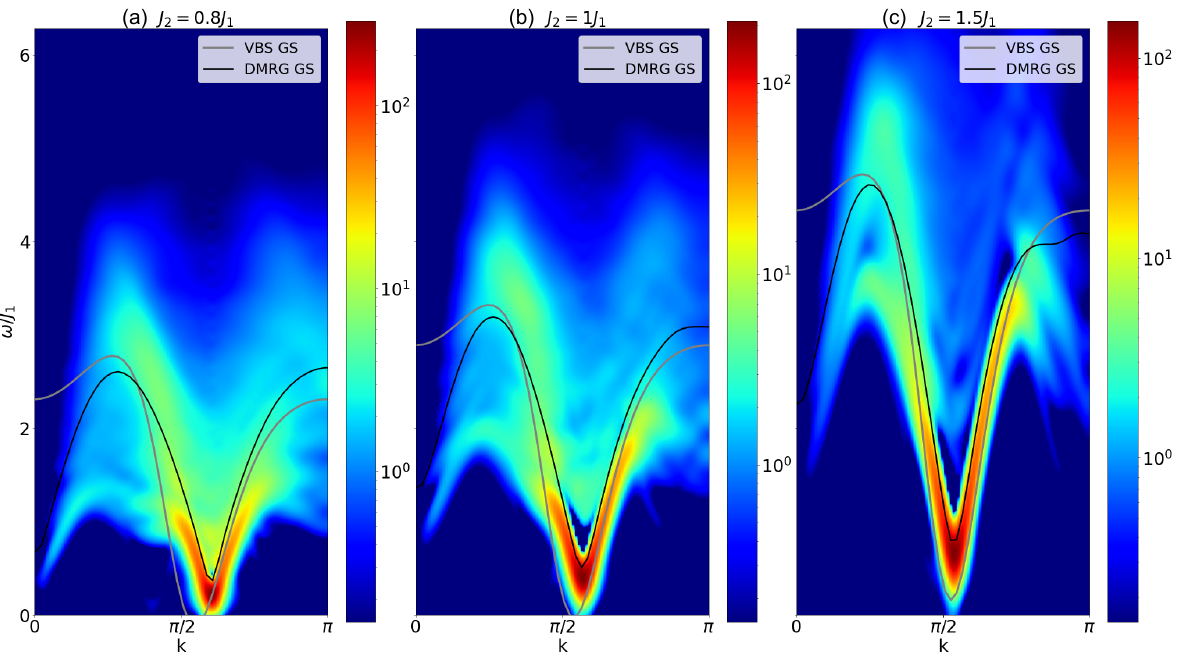}
\caption{ DSF of the $J_1 - J_2$ spin-1 chain in the limit of large $J_2$. The magnon SMA dispersion calculated from the NNN VBS state and the groundstate obtained from DMRG has been plotted on the DSF.}
\label{Fig11_nnnaklt}
\end{figure}

In the large $J_2$ limit,  the spin-1 chain can be treated as two nearly decoupled spin-1 chains. Since the unit cell length is doubled, the Brillouin zone size is halved. The dispersion of the  magnon excitation in the DSF is very similar to that in the Haldane phase,  but with reduced periodicity of $\pi$ (see Fig. \ref{Fig11_nnnaklt}). The SMA calculations for the magnon dispersion in NNN VBS state and the ground state obtained from DMRG  have been plotted with good agreement to the magnon mode found in DSF around $k=\pi$. As $J_2$ increases, the NNN VBS state magnon excitation approaches  the magnon excitation over the DMRG calculated ground state. A similar discussion on the bound states of magnons can be expected for this phase, however it is difficult to identify the modes since the DSF we computed is still for $J_2$ values  comparable to $J_1$. In order to follow a systematic analysis of the spectra one should start from the limit $J_2 \gg J_1$,  where the magnon mode and its associated continua can be determined unambiguously. Decreasing the interaction strength $J_2$ from large values to the phase transition point, one would in practice follow the evolution of the  spectra and identify the bound state modes separating from the continua. This is left for future studies.

\section{Conclusion}
\label{Conclusion}
	
	The $J_1-J_2$ spin-1 chain is a nice platform to study exotic phenomena as its spectrum at various values of $J_2$ realizes  -  (i) fractionalization of quasi-particles, (ii) onset of incommensurability, and (iii) confinement - deconfinement of the fractionalized quasi-particles. This is in contrast to spin-$1/2$ case, for which one needs to add explicit dimerization term to study the last phenomena\cite{uhrig1999unified, lavarelo2014spinon}. Our numerical study describes the dispersion relation of the magnon mode and the onset of incommensurability in the dispersion. By constructing the multi-magnon continua, one is able to distinguish between bound states and resonances. The smooth evolution from magnon picture to spinons in a confining potential, only to be deconfined at the phase transition, is demonstrated by the DSF at various $J_2$ coupling strengths. Although the spinon dispersion had been determined using a quasi-particle ansatz \cite{vanderstraeten2020spinon}, we showed that the continuum constructed from such a spinon-dispersion does not explain the spectra seen in the numerical DSF at the phase transition. Assisted with the form of the spinon dispersion in the low-energy limit, our numerical DSF and the SMA, we extracted the single spinon dispersion throughout the Brillouin zone. We found a simple picture of competing tight-binding hoppings to describe the onset of incommensurability in the spinon dispersion. The DSF also contains spectral features associated with the confinement of the spinons and we identified the bunching of spectral features at the incommensurate momentum corresponding to the minima of the spectra to be the boundstates of spinons. We also characterised the gap between the excited states as $J_2$ is varied in the vicinity of the phase transition.\par
	The simple $J_1-J_2$ spin chain is a very realistic model and it can be realized in nature as zig-zag chains. The spin-1/2 $J_1-J_2$ chain is realized in copper-based compound $\mathrm{Cu}\mathrm{Ge}\mathrm{O}_3$ (although it undergoes a spin-Peierls transition) and its spectrum has been studied by INS experiments \cite{ain1997double}. Since the DSF is proportional to the differential scattering cross-section measured in the INS experiments, the reported spectra can be experimentally checked for compounds realizing the spin-1 zigzag chain. With appropriate crystal field splitting, $\mathrm{Ni}^{2+}$-based compound (such as $\mathrm{Ni} {\left(\mathrm{C}_2\mathrm{H}_8\mathrm{N}_2\right)}_2\mathrm{N}\mathrm{O}_2\mathrm{Cl}\mathrm{O}_4$(NENP) \cite{ma1992dominance} and $\mathrm{Cs}\mathrm{Ni}\mathrm{Cl}_3$ \cite{hagiwara1990observation, kenzelmann2002properties}) can realize nearest-neighbour anti-ferromagnetic spin-1 chains although nickel based compounds usually also give rise to on-site anisotropy terms. In that case, the present results will be still very useful to understand the spectral functions. \par
	
	The same numerical and analytical tools can be useful in studying the spectra of the $J_1-J_2$ spin chain for larger spin values. The groundstate phase diagrams for spin-$3/2$ and spin-$5/2$ have been explored in the literature \cite{roth1998,chepiga2020floating,chepiga2022from} and different phases are realized in these chains as compared to spin-$1/2$ and spin-$1$. As $J_2$ is tuned, the system undergoes a phase transition from a gapless phase to a partially dimerized phase where the incommensurate real space-spin correlations set in at $J_2\approx 0.38J_1$ ($J_2\approx 0.33 J_1$) for spin-$3/2$ ($5/2$) chain. In contrast to the spin-1/2 phase diagram, the system transitions into a floating phase where the spectral gap remains closed throughout the phase and  opens slowly at larger values of $J_2$ when the system transitions into the fully dimerized phase. Since different ground states are realized in each phase, one can explore the nature of the lowest lying excitations in the partially dimerized and floating phases and interpret the spectra with these building blocks. The $J_1-J_2$ spin-$3/2$ and spin-$5/2$ chains are also very realistic models to exist in nature - particularly in transition-metal based compounds \cite{boya2021magnetic}. Therefore, a spectral analysis of such chains can reveal exciting possibilities of exotic phenomena being realized in real compounds via neutron scattering experiments.

\begin{acknowledgements}
We thank Natalia Chepiga, Ivo Maceira, Samuel Nyckees, and Laurens Vanderstraeten for helpful discussions. A.S acknowledges the support of Swiss Government Excellence Scholarship FCS grant(2021.0414) during the work.  This work has been supported by the
Swiss National Science Foundation Grant No. 212082. Simulations were performed on Jed cluster managed by SCITAS team at EPFL, Lausanne.
			
\end{acknowledgements}

\bibliography{manuscript_version7.bbl}
		
\section*{APPENDIX}
		
\appendix

\section{VBS ground state energies}

\begin{figure}
\centering
\includegraphics[width=9.5cm, height = 9.5cm, keepaspectratio]{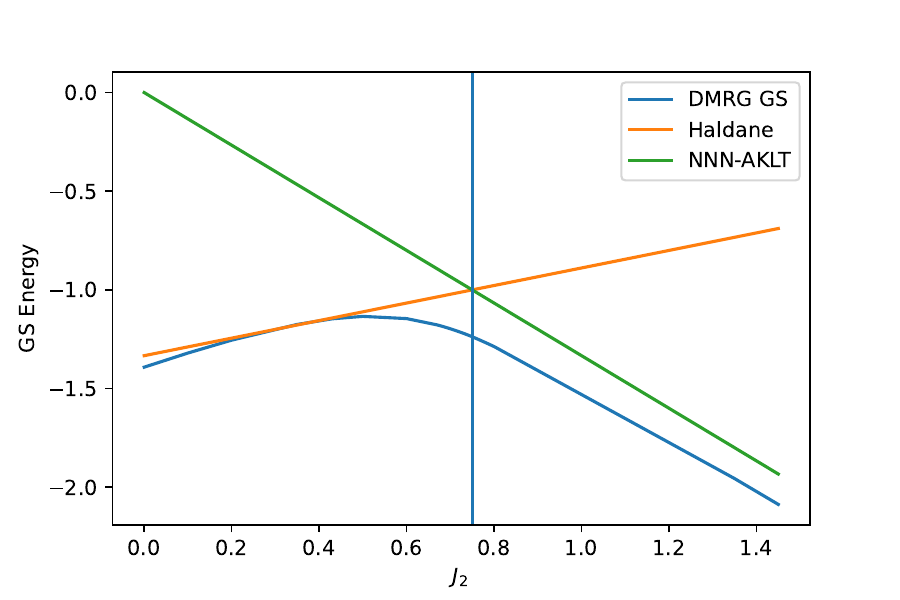}
\caption{ Variation of the AKLT and the NNN AKLT state's energies across $J_2$. The AKLT state energies vary linearly with $J_2$  with a slope of $4/9$ while the NNN AKLT state energies vary with a negative slope of $4/3$. The intersection point of the energies is marked by a blue line, which is close to the first-order phase transition.}
\label{Fig_appendix_vbsenergy}
\end{figure}

The ground state of the $J_1 - J_2$ spin-1 chain is regularly pictured as a VBS ground state in the main text to explain the main excitations. In. Fig. \ref{Fig_appendix_vbsenergy}, we compare the ground state energies obtained from MPS calculations with the energies of the VBS state as $J_2$ is varied. The VBS states were encoded as MPS states and the energies were found by contraction with the Matrix Product Operator (MPO) representing the $J_1 - J_2$ spin-1 Hamiltonian. The AKLT state and NNN AKLT state energies vary linearly with $J_2$ and intersect at $J_2 = 0.75 J_1$ indicating the position of the first-order phase transition. Since the phase transition is a {\it weakly} first-order phase transition, the ground state energies with $J_2$ look smooth across the phase transition. The characteristic kink in the groundstate energies for first-order transition is too small to resolve with our MPS parameters. However, measuring the minimum of the spectral gap, we find that the phase transition takes place at $J_2 = 0.76 J_1$ which is close to the location of the intersection of the AKLT and the NNN AKLT ground state energies. 

\section{Encoding the domain wall state as a MPS}
\label{spinoncalc}

We revisit the procedure to encode an AKLT and NNN AKLT state as a MPS and describe the way to construct a domain wall state as a MPS \cite{kolezhuk1997variational, schollwock2011density}.  The spin-$1$ degrees of freedom  at each site can be seen to be a symmetric combination of two spin-$1/2$ states - $|\uparrow\rangle$ and $|\downarrow\rangle$. The symmetrized combination of two spin-$1/2$s leads to three triplet states with total magnetization $ 1, ~0, ~-1$ at each site and therefore one can define a projection operator $\hat{P}_T$ on the full chain to be 

\begin{eqnarray}
\hat{P}_T = \sum_{\lbrace \sigma \rbrace}\sum_{\lbrace a\rbrace, \lbrace b\rbrace}\prod_{i}P^{\sigma_i}_{a_i, b_i}|\sigma_i\rangle \langle a_i|\langle b_i|
\end{eqnarray}
where, $\sigma_i$ is the physical spin-1 degree of freedom. The matrices $P^{\sigma_i}_{a_i, b_i}$ maps the two spin-$1/2$ s (denoted by $a_i$ and $b_i$ ) on $i$-th site to the spin-1 degree of freedom. The full projector is obtained by summing over the spin-$1/2$ and spin-$1$ states on all the sites of the chain.

\begin{enumerate}
\item {\it Preparing the AKLT state}:
The adjacent spin-$1/2$s on the neighboring physical sites can be combined anti-symmetrically as a singlet state. Because of the open boundary condition, the spin-$1/2$s at the physical sites on the edges of the chain are free. This leads to a product state consisting of $2N$ - spin-$1/2$ sites as 
\begin{eqnarray}
|\Sigma_{T}\rangle = \sum_{\lbrace a \rbrace}\sum_{\lbrace b \rbrace}\left(\prod_{i =1}^{N-1} \Sigma_{b_{i},a_{i+1}}|b_i\rangle | a_{i+1}\rangle\right) |a_1\rangle |b_{N}\rangle,
\end{eqnarray}
 where the sum over all spin-$1/2$s including the edge spin-$1/2$s results in a single wavefunction with the edge spins polarized along $x$-axis. This choice of the boundary condition does not affect any result reported in the paper.

The AKLT state is then obtained by acting the full projector on the product singlet state :
\begin{eqnarray}
|\psi_{\mathrm{AKLT}}\rangle &=&  \sum_{\lbrace \sigma \rbrace, \lbrace a\rbrace,\lbrace b\rbrace}  \prod_{i} P^{\sigma_i}_{a_i,b_i}\Sigma_{b_i, a_{i+1}}\prod_{i}|\sigma_i\rangle\\
&=&\sum_{\lbrace \sigma \rbrace, \lbrace a\rbrace,\lbrace b\rbrace}  \left(\prod_{i = 1}^{i = N-1} M^{\sigma_i}\right)P^{\sigma_{N}}_{a_N, b_N}\prod_{i}|\sigma_i\rangle\nonumber,
 \end{eqnarray}
where $M^{\sigma_i}$ is the product of the local projector matrix (denoted by $P^{\sigma_i}_{a_i,b_i}$) and the singlet operator between neighbouring sites (denoted by $\Sigma_{b_i, a_{i+1}}$). 
\item {\it Preparing the NNN AKLT state}:  The singlet bond state is obtained by combining anti-symmetrically the nearest spin-$1/2$s on the next-nearest-neighbouring physical sites. Because of the open boundary condition, one of the spin-$1/2$s at the physical sites on the edges of the chain are combined with the nearest neighbour site's spin-$1/2$s to form singlets. Therefore, the product state is given by :
\begin{eqnarray}
|\Sigma_{T, \mathrm{NNN}}\rangle &=& \sum_{\lbrace a\rbrace}\sum_{\lbrace b\rbrace} \left(\prod_{i = 1}^{N-2}\Sigma_{b_{i},a_{i+2}}|b_i\rangle | a_{i+2}\rangle\right)\times\nonumber\\
&&\Sigma_{a_1, a_2}|a_1\rangle |a_2\rangle\Sigma_{b_{N-1}, b_{N}}|b_{N-1}\rangle |b_{N}\rangle
\end{eqnarray} 
The NNN AKLT state is obtained by acting the full projector on the next nearest neighbour singlet state:
\begin{eqnarray}
&&|\psi_{\mathrm{NNN AKLT}}\rangle \nonumber\\
&=&\sum_{\lbrace \sigma \rbrace, \lbrace a\rbrace,\lbrace b\rbrace} \Sigma_{a_{1}, a_{2}}\left(\prod_{i=1}^{N-2} P^{\sigma_i}_{a_i,b_i}\Sigma_{b_i, a_{i+2}}\right)\times\nonumber\\
&&P^{\sigma_N}_{a_{N-1},b_{N-1}}\Sigma_{b_{N-1}, b_{N}}P^{\sigma_{N}}_{a_N,b_N}\prod_{\otimes i}|\sigma_i\rangle\nonumber\\
&=&\sum_{\lbrace \sigma \rbrace, \lbrace a\rbrace,\lbrace b\rbrace} \Sigma_{a_{1}, a_{2}}\left(\prod_{i=1}^{N-2} \tilde{M}^{\sigma_i}\right)\times\nonumber\\
&&M^{\sigma_{N-1}}_{a_{N-1}, b_{N}}P^{\sigma_N}_{a_N, b_N}\prod_{\otimes i}|\sigma_i\rangle,
\end{eqnarray}
where $\tilde{M}^{\sigma_i}$ is the product of the local projector matrix (denoted by $P^{\sigma_i}_{a_i,b_i}$) and the singlet operator between next-nearest neighbouring sites (denoted by $\Sigma_{b_i, a_{i+2}}$). The tensor $M^{\sigma_{N-1}}_{a_{N-1}, b_{N}}$ is the product of the projector at the $N-1$ th site with the nearest neighbour singlet matrix.
\end{enumerate}

Generating a state encoding a single domain wall between the AKLT and the NNN AKLT state at the $j$-th site involves constructing the AKLT state up to $j$-th site and then continuing the construction with the NNN AKLT state from $j+1$- th site until the end of the chain. This generates the spinon-state shown in the Fig.\ref{Fig7_spinon_fitting}b in the main text. It leaves one spin-$1/2$ degree of freedom at the $j$-th site free. 
\begin{eqnarray}
|\Omega_j\rangle &=&  \sum_{\lbrace \sigma \rbrace, \lbrace a\rbrace,\lbrace b\rbrace}\left(\prod_{i = 1}^{i =j-1}M^{\sigma_i}\right) P^{\sigma_{j}}_{a_j, b_j} \Sigma_{a_{j+1} a_{j+2}}\times\nonumber\\
&&\left(\prod_{i = j+1}^{N-2}\tilde{M}^{\sigma_i}\right)M^{\sigma_{N-1}}_{a_{N-1}, b_{N}}P^{\sigma_N}_{a_N, b_N}\prod_{\otimes i}|\sigma_i\rangle
\label{SMA_dispersion_spinon_state}
\end{eqnarray}
By defining a singlet state between the $i$-th and the $j$-th sites and a projector on $i$-th site according to
\begin{eqnarray}
|S_{i,j}\rangle &=& \sum_{b_i, a_j}\Sigma_{i,j}|b_i\rangle |a_j\rangle\\
\hat{P}^{\sigma_i} &=& \sum_{a_i,b_i} P^{\sigma_i}_{a_i, b_i}|\sigma_i\rangle\langle a_i|\langle b_i|,
\end{eqnarray}
Eq. \ref{SMA_dispersion_spinon_state} can be simplified as
\begin{eqnarray}
|\Omega_j\rangle &=& \sum_{\lbrace \sigma\rbrace}\left(\prod_{i=1}^{N} \hat{P}^{\sigma_i}\right)|a_1\rangle\left( \prod_{ i=1}^{j-1} |S_{i,i+1}\rangle\right)|b_j\rangle |S_{j+1,j+2}\rangle \times\nonumber \\
&&\left(\prod_{ i=j+1}^{N-2} |S_{i,i+2}\rangle\right) |S_{N-1,N}\rangle,
\end{eqnarray}
where $|b_j\rangle$ is the unpaired spin-$1/2$ state at the $j$-th site and $|a_1\rangle$ is the free spin-$1/2$ state at the edge of the chain. The sum over all the spin-$1/2$ degrees of freedom results in a single wavefunction where the unpaired spin-$1/2$ at site $j$ and the free edge spin-$1/2$ are $x$-polarized before projection. This does not affect any results reported in the paper.     
Using the encoding of the domain wall state as a MPS, one can find the dispersion relation using the SMA analysis \cite{lavarelo2014spinon} as described in the main text. The state we generate is not normalized, but the normalization factor of the state cancels out from numerator and denominator, so the dispersion relation is unaffected.  

		
\section{Propagation of spin-spin correlations} 
	
	The spread of the spin-spin correlations with time follows a cone-like structure, and one can extract the velocity of the correlation spread wave-front by tracking the maximum of the spin-spin correlation function at a given time slice. The velocity was obtained by determining the slope of the position of the maximum of the correlation function in the chain as a function of time. The spread of the correlations at early times are more trustworthy as compared to later times due to the accumulation of Trotter and truncation errors over time. Therefore, we considered the spin-spin correlations for the  first 20 - 40 time steps to extract the velocity of the spread of correlations.\par
\begin{figure}
\centering
\includegraphics[width=10cm, height = 10cm, keepaspectratio]{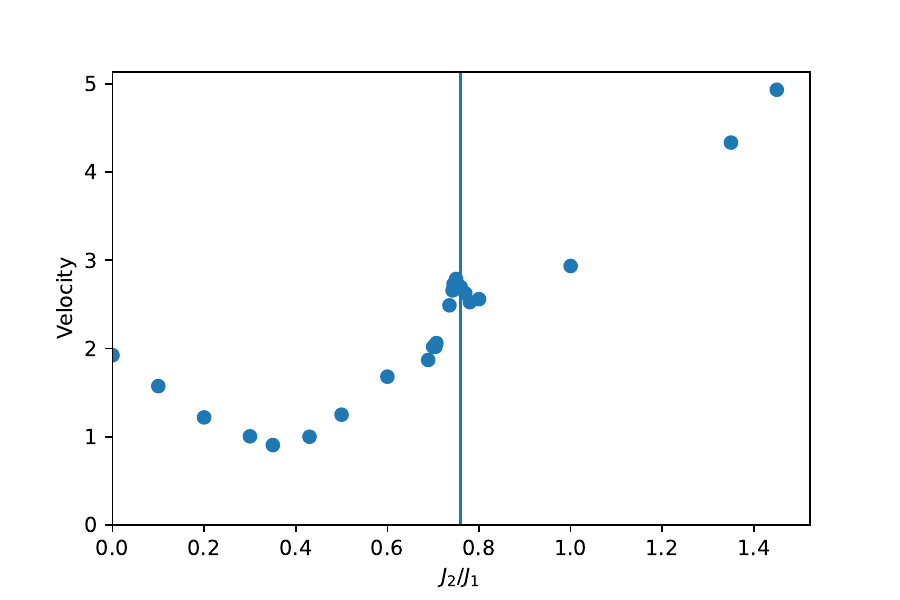}

\caption{ Velocities of the spread of the spin-spin correlation functions  versus $J_2$ interactions. The velocity decreases linearly up to the disorder point beyond which it increases linearly, with a kink appearing near the transition point (denoted by a vertical line).}
\label{Fig_appendix_velocities}
\end{figure}

  
	This velocity (denoted by $v$) is compared across $J_2$ values (see Fig.\ref{Fig_appendix_velocities}). It decreases linearly as $J_2$ increases until the disorder point. The velocity then increases linearly for all further $J_2$ values, apart from a sharp kink observed in the plot near the transition point. Numerically, when determining the DSF of a finite chain of $N$-sites, we considered spin-spin correlations calculated up to the final time $t_f \sim N/2v$ in order to be free from spectral artefacts induced by the boundaries of the chain.
		
\section{Determining the amplitude of the spinon dispersion}
\label{spinonheight}

\begin{figure}
\centering		
\includegraphics[width = 8.5cm , height =8.5cm , keepaspectratio]{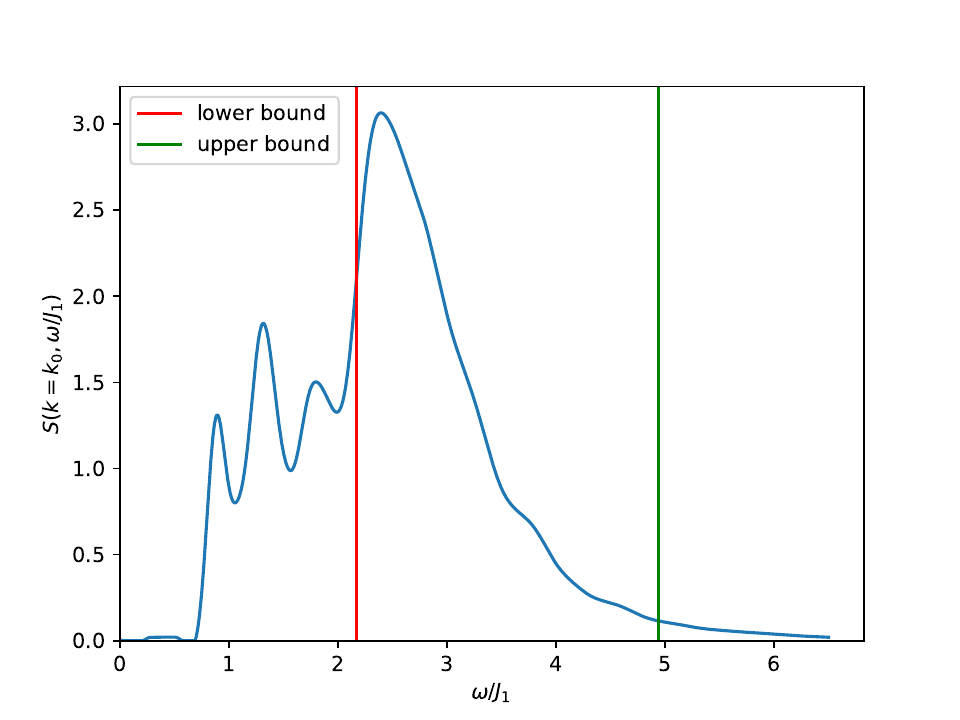}			
\caption{ Comparing the lower (shown in red) and upper boundaries (shown in green) of the spinon-continua generated from the spinon dispersion at $k=\pi$ with an amplitude of $\sim2.48 J_1$ with the section cut of the DSF at the phase transition point ($J_2=0.76J_1$). The lower boundary of the continua lies on the inflection point between the highest spectral peak and the local minima slightly lower in energy.} 
\label{Fig_appendix_upperspinon}
\end{figure}

In the main text, we describe the way to arrive at the most reasonable estimate of the spinon dispersion relation, for which we needed to determine the amplitude of the dispersion relation. This information can be inferred directly from the numerically computed DSF at the phase transition point  $J_2 = 0.76J_1$ by finding the lower bound of the two-spinon continuum in the neighbourhood of $k = \pi$. The most plausible location for the lower bound is the inflection point between the highest spectral peak and the local minima slightly lower in energy.\par

Note that identifying a single unambiguous inflection point is difficult. This is because there are multiple modes below the spinon continuum at the phase transition point which are the magnons or the multi-magnon bound state modes arising from excitations on top of VBS or NNN VBS like states. These modes appear as well-identifiable spectral peaks in the section cuts of the DSF in the neighbourhood of $k = \pi$, but as one moves further away from $k=\pi$, these modes enter the two-spinon continuum leading to broadening of the spectral peak and resulting in multiple inflection points. Fortunately, at $k = \pi$, there are only three well defined spectral peaks corresponding to individual modes, and we were able to identify the inflection point clearly by calculating the zeroes of the double-derivative of the spectral weights between the local minima following the third peak and the highest peak.  
 
Then we fit the energy of this inflection point with the lower bound of the continuum from the spinon dispersion at $k=\pi$, using the amplitude of the spinon dispersion as a fitting parameter. This leads to a dispersion amplitude of $ \sim 2.48 J_1$, hence to a maximum of the two spinon continuum just below $5 J_1$, as shown in Fig. \ref{Fig_appendix_upperspinon}.

\end{document}